\documentclass[12pt]{article} 
\usepackage[sectionbib]{natbib}
\usepackage{array,epsfig,rotating}
\usepackage[]{hyperref}  
\usepackage{sectsty, secdot}
\sectionfont{\fontsize{12}{14pt plus.8pt minus .6pt}\selectfont}
\renewcommand{\theequation}{\thesection\arabic{equation}}
\subsectionfont{\fontsize{12}{14pt plus.8pt minus .6pt}\selectfont}

\usepackage{amsmath}
\usepackage{amssymb}
\usepackage{amsfonts}
\usepackage{multirow}
\usepackage{natbib}
\usepackage{amsthm}
\usepackage{graphicx}
\usepackage{booktabs}
\usepackage[tight]{subfigure}
\usepackage{float}
\usepackage{arydshln}
\usepackage{adjustbox}
\usepackage{blindtext}
\usepackage{float}
\usepackage[normalem]{ulem}
\usepackage{tikz}
\usepackage{cancel}
\usepackage{lscape}
\usepackage{rotating}

\usepackage{etoolbox}
\usepackage[doublespacing]{setspace}
\AtBeginEnvironment{tabular}{\singlespacing}

\newcommand{\be}{\begin{equation}}
\newcommand{\ee}{\end{equation}}
\newcommand{\bea}{\begin{eqnarray}}
\newcommand{\eea}{\end{eqnarray}}
\newcommand{\mbf}[1]{\mathbf{#1}}
\newcommand{\mbs}[1]{\boldsymbol{#1}}
\newcommand{\mcal}[1]{\mathcal{#1}}

\newcommand{\by}{\mbf{y}}

\newcommand{\ba}{\mbf{a}}
\newcommand{\bb}{\mbf{b}}

\newcommand{\br}{\mbf{r}}
\newcommand{\bs}{\mbf{s}}
\newcommand{\bh}{\mbf{h}}

\newcommand{\bt}{\mbf{t}}

\newcommand{\bv}{\mbf{v}}
\newcommand{\bx}{\mbf{x}}
\newcommand{\bw}{\mbf{w}}

\newcommand{\bz}{\mbf{z}}

\newcommand{\bB}{\mbf{B}}
\newcommand{\bC}{\mbf{C}}

\newcommand{\tbC}{\tilde{\bC}}
\newcommand{\bF}{\mbf{F}}

\newcommand{\bV}{\mbf{V}}
\newcommand{\bX}{\mbf{X}}

\newcommand{\bW}{\mbf{W}}

\newcommand{\bSigma}{\mbs{\Sigma}}
\newcommand{\bGamma}{\mbs{\Gamma}}
\newcommand{\bUps}{\mbs{\Upsilon}}
\newcommand{\bPsi}{\mbs{\Psi}}

\newcommand{\bP}{\mbf{P}}

\newcommand{\bpsi}{\mbs{\psi}}
\newcommand{\bphi}{\mbs{\phi}}
\newcommand{\bmu}{\mbs{\mu}}
\newcommand{\bbeta}{\mbs{\beta}}

\newcommand{\bchi}{\mbs{\chi}}
\newcommand{\blambda}{\mbs{\lambda}}
\newcommand{\bLambda}{\mbs{\Lambda}}

\newcommand{\bepsilon}{{\mbs{\varepsilon}}}
\newcommand{\bgamma}{{\mbs{\gamma}}}
\newcommand{\btheta}{{\mbs{\theta}}}

\newcommand{\0}{\mbs{0}}

\newcommand{\lrp}[1]{\left(#1\right)}
\newcommand{\lrb}[1]{\left\{#1\right\}}

\newcommand{\ben}{\begin{equation*}}
\newcommand{\een}{\end{equation*}}
\newcommand{\bean}{\begin{eqnarray*}}
\newcommand{\eean}{\end{eqnarray*}}
\newcommand{\bsm}{\begin{smallmatrix}}
\newcommand{\esm}{\end{smallmatrix}}
\newcommand{\bmat}{\begin{matrix}}
\newcommand{\emat}{\end{matrix}}
\newcommand{\tI}{\text{I}}
\newcommand{\tN}{\text{N}}

\newcommand{\parent}{\mcal{P}}

\newcommand{\nngpw}[4]{\text{NNGP}_{#1}^{#2}{\lrp{0,\tilde{\mcal{C}}_{#2}(\cdot\given \phi_{#3}^{#4})}}}
\newcommand{\gpw}[4]{\text{GP}_{#1}^{#2}{\lrp{0,\mcal{C}_{#2}(\cdot\given \phi_{#3}^{#4})}}}

\newcommand{\refset}{\mcal{T}}

\newcommand{\oset}{\mcal{T}}

\newcommand{\given}{\,|\,}

\theoremstyle{definition}

\begin{document}


\markright{ \hbox{\footnotesize\rm Statistica Sinica
}\hfill\\[-13pt]
\hbox{\footnotesize\rm
}\hfill }

\markboth{\hfill{\footnotesize\rm Daniel Taylor-Rodriguez  AND Andrew Finley AND Abhirup Datta AND Sudipto Banerjee} \hfill}
{\hfill {\footnotesize\rm Spatial Factor Models for High-Dimensional and Large Spatial Data: An Application in Forest Variable Mapping} \hfill}

\renewcommand{\thefootnote}{}
$\ $\par


\fontsize{12}{14pt plus.8pt minus .6pt}\selectfont \vspace{0.8pc}
\centerline{\large\bf Spatial Factor Models for High-Dimensional and }
\centerline{\large\bf  Large Spatial Data: An Application in Forest Variable Mapping}
\vspace{.4cm} \centerline{Daniel Taylor-Rodriguez$^1$, Andrew O. Finley$^{2,*}$, Abhirup Datta$^3$, Chad Babcock$^4$,}
\centerline{Hans-Erik Andersen$^5$, Bruce D. Cook$^6$, Douglas C. Morton$^6$, Sudipto Banerjee$^7$} \vspace{.4cm} 
\centerline{\it$^1$Portland State University, $^2$Michigan State University, $^3$Johns Hopkins University,}
\centerline{\it$^4$University of Washington, $^5$United States Forest Service,}
\centerline{\it$^6$National Aeronautics and Space Administration,}
\centerline{\it $^7$University of California Los Angeles, $^*$ Corresponding}\vspace{.55cm} \fontsize{9}{11.5pt plus.8pt minus
.6pt}\selectfont


\begin{quotation}
\noindent {\it Abstract:} Gathering information about forest variables is an expensive and arduous activity.  As such, directly collecting the data required to produce high-resolution maps over large spatial domains is infeasible. Next generation collection initiatives of remotely sensed Light Detection and Ranging (LiDAR) data are specifically aimed at producing complete-coverage maps over large spatial domains. Given that LiDAR data and forest characteristics are often strongly correlated, it is possible to make use of the former to model, predict, and map forest variables over regions of interest.  This entails dealing with the high-dimensional ($\sim$$10^2$) spatially dependent LiDAR outcomes over a large number of locations ($\sim$$10^5-10^6$).  With this in mind, we develop the Spatial Factor Nearest Neighbor Gaussian Process (SF-NNGP) model, and embed it in a two-stage approach that connects the spatial structure found in LiDAR signals with forest variables. We provide a simulation experiment that demonstrates inferential and predictive performance of the SF-NNGP, and use the two-stage modeling strategy to generate complete-coverage maps of forest variables with associated uncertainty over a large region of boreal forests in interior Alaska. 

\vspace{9pt}
\par
\end{quotation}\par

\def\thefigure{\arabic{figure}}
\def\thetable{\arabic{table}}

\renewcommand{\theequation}{\thesection.\arabic{equation}}

\fontsize{12}{14pt plus.8pt minus .6pt}\selectfont

\setcounter{section}{0} 
\setcounter{equation}{0} 

\section{Introduction}\label{sec:intro}

Strong relationships between remotely sensed Light Detection and Ranging (LiDAR) data and forest variables have been documented in the literature \citep{asner2009, babcock2013, naesset2011}. When used in forested settings, LiDAR data provide a high-dimensional signal that characterizes the vertical structure of the forest canopy at point-referenced locations. Traditionally LiDAR data acquisition campaigns have sought complete coverage at a high spatial resolution over relatively small spatial domains---resulting in a fine grid of point-referenced LiDAR signals. In such settings, the link between LiDAR data and forest variable measurements on sparsely sampled forest inventory plots has been exploited to create high resolution complete-coverage predictive maps of the forest variables. Commonly this link is established by first extracting relevant features of the high-dimensional LiDAR signals through a dimension reduction step \citep{babcock2015, junttila2017}, then using the LiDAR features as predictors in a regression model to explain variability in spatially coinciding forest variable outcomes. The model is then applied to predict the forest outcomes at all locations across the domain where LiDAR signals have been observed.  

Considerably more ambitious next generation LiDAR collection initiatives, such as ICESAT-2 \citep{ICESAT2015}, Global Ecosystem Dynamics Investigation LiDAR (GEDI) \citep{GEDI2014}, and NASA Goddard's LiDAR, Hyper-Spectral, and Thermal imager (G-LiHT) \citep{GLIHT2016}, seek to quantify and map forest variables over vast spatial extents. To fulfill their goals in a cost effective manner, these data gathering programs do not collect LiDAR data over the entire domain, but rather sparsely sample locations across the domain extent and over forest inventory plots (i.e., where forest variables have been measured).  While generating complete-coverage high resolution maps of forest outcomes remains the primary intended use for these data, there is also interest in creating maps of LiDAR data over non-sampled locations, and assessing spatial dependence within and among LiDAR signals.  

Our motivating application focuses on forest variable prediction and mapping in the boreal forests of interior Alaska using sparsely sampled LiDAR and forest variable measurements. Within these regions, acquiring complete coverage LiDAR is cost prohibitive \citep{andersen2011, bolton2013, nelson2012}. Because complete coverage maps of forest variables (and perhaps LiDAR signals) is still the goal, the information in the sparsely sampled LiDAR must be leveraged to inform forest variable prediction. One attractive solution is to move the LiDAR predictor variables to the left hand side of the regression and model them jointly with the forest outcomes. When the number of LiDAR and forest variables is small, such joint models are possible via linear models of coregionalization, see, e.g., \cite{babcock2017} and \cite{finley2014b}.Alternatively, if the LiDAR signal is high-dimensional but observed at a small number of locations reduced rank models can be employed. For example, \cite{banerjee2008gaussian}, \cite{Ren2013}, and \cite{Finley2017a} applied a reduced rank \emph{predictive process} modeling strategy to analyze similar high-dimensional data. However, such approaches that employ a reduced rank representation of the desired spatial process cannot scale to datasets with tens of thousands of locations and can yield poor predictive performance \citep{stein2014limitations}.

Models able to handle high-dimensional signals observed over a large number of locations and capable of estimating within and among location dependence structures are needed. Recent modeling developments reviewed in \cite{heaton2017} and \cite{banerjee2017} highlight several options for robust and practical approximation of univariate Gaussian Process (GP) models. A subset of these models can be easily extended to accommodate relatively small multivariate response vectors (5 or less) see, e.g., \citep{Datta2016c}; nevertheless, for our particular application we require an approach that can cope with both the high-dimensional LiDAR measurements, $\sim$50 outcomes at a location, while making use of the large collection of observed locations.  

The Nearest Neighbor Gaussian Process (NNGP) developed in \cite{Datta2016c}, \cite{Datta2016}, and \cite{Datta2016a} can be used with a massive number of locations as its scalability is not mediated by the number of observed locations, but rather by the size of the nearest neighbor sets considered---a quality that yields minimal storage and computational requirements. These models belong to the class of methods that induce sparsity on the spatial precision matrix, and exploits the natural representation of sparsity provided by graphical models \citep{lauritzen96,murphy2012} to build a sparse GP that accurately approximates the original dense GP. 

To tackle the high-dimensional LiDAR dataset, we develop a Bayesian NNGP spatial factor model (SFM), referred to as the SF-NNGP. Following \citet{christensen2002latent,hogan2004bayesian,Ren2013} the SFM structure enables approximating the dependence between multivariate (spatially dependent) outcomes through a lower-dimensional set of spatial factors, alleviating the difficulty of dealing directly with high-dimensional outcomes. 
The SF-NNGP allows us to model and map the LiDAR signals on both observed and unobserved locations, and, conditioning on the LiDAR spatial signatures, we can likewise map the forest variables over the entire spatial domain of interest. Furthermore, using a Bayesian approach for model fitting enables us to equip the derived estimates and predictions with associated measures of uncertainty; an essential requirement of many high-profile initiatives. Our methods are fully implemented in \textsf{C}$++$, using \textsf{BLAS} \citep{blackford2001,zhang13} to leverage efficient multi-processor matrix operations and \textsf{openMP} \citep{openmp98} to improve key steps of the algorithm through parallelization. \emph{Code and reproducible results will be provided via a GitHub site prior to publication.}

The structure for the remainder of document is as follows.  Section \ref{sec:data} introduces the Bonanza Creek dataset. In Section \ref{sec:model} we formulate the proposed hierarchical Bayesian modeling strategy.  Section \ref{sec:sims} presents the analysis of a synthetic dataset to validate the performance of the SF-NNGP model. Using the available LiDAR and forest inventory data, in Section \ref{sec:lidar} we develop and validate a predictive model for forest  variables.  We close by providing some insights, recommendations and future directions in Section \ref{sec:discussion}.

\section{Data Description}\label{sec:data}

\begin{figure}[H]
\centering 
\includegraphics[scale=1,trim={0cm 3.5cm 0.3cm 3.5cm},clip]{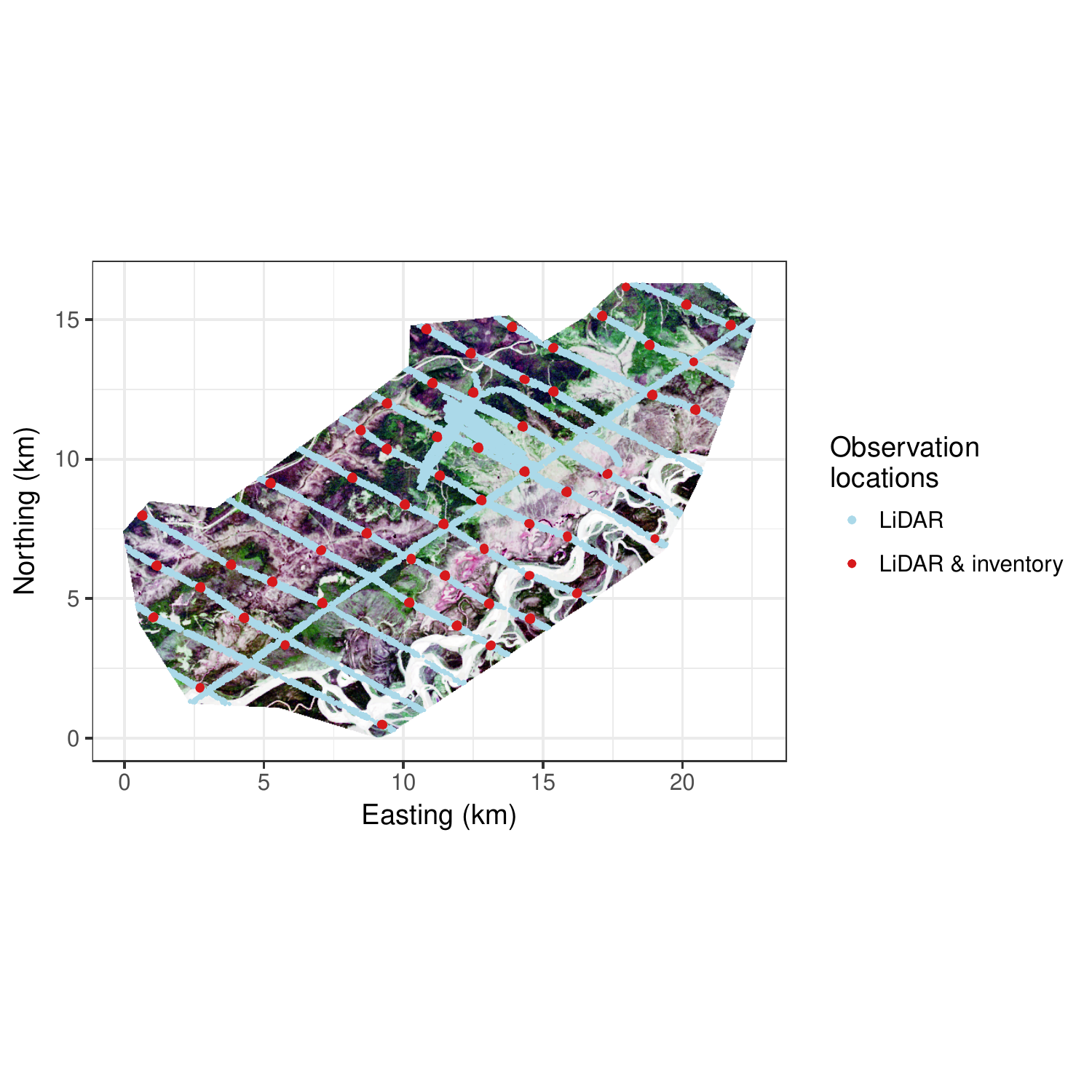}
\caption{Bonanza Creek Experimental Forest extent with color enhanced Landsat image and locations where the LiDAR signals were measured (\emph{LiDAR} in the legend) and locations where both LiDAR signals and forest variables were measured (\emph{LiDAR \& inventory} in the legend).}\label{fig:bc-map}
\end{figure}

The Bonanza Creek Experimental Forest (BCEF) is a Long-Term Ecological Research (LTER) site consisting of vegetation and landforms typical of interior Alaska. The BCEF is 21,000 ha and includes a section of the Tanana River floodplain along the southeastern borders \citep{BCEF2016}. Figure~\ref{fig:bc-map} shows the location and extent of the BCEF data detailed in this section. 

Forest variables were collected on 197 plots in 2014 using the USDA Forest Service Forest Inventory and Analysis Program protocol \citep{bechtold2005}. We consider three forest variables commonly used by forest professionals to make management decisions: above-ground biomass (AGB); tree density (TD); basal area (BA). AGB for individual trees was estimated using the Component Ratio Method described in \citet{woodall2015}. TD for a plot is expressed in thousands of trees per ha. BA for a plot is the sum of individual trees' cross-sectional areas in $m^2$ at breast height scaled to a per ha basis. 

In the summer 2014 LiDAR data were collected using a flight-line strip sampling approach with NASA Goddard's G-LiHT sensor \citep{cook2013}, which is a portable multi-sensor system that accurately characterizes complex terrain and vertical distribution of canopy elements \citep{jakubowski2013,white2013}. Point cloud information was summarized to a 13$\times$13 m grid cell size to approximate field plot areas. Over each grid cell, psuedo-waveforms were generated by calculating LiDAR return count densities for .5 m height bins between 0 and 28.5 m (i.e., 57 LiDAR outcomes per location). LiDAR return count density for height bin $l$ is defined as the number of returns in height bin $l$ divided by the total number of LiDAR returns over the grid cell. Identical LiDAR psuedo-waveforms were obtained using point clouds extracted over each field plot.
G-LiHT data for the study area are available online at \url{https://gliht.gsfc.nasa.gov}. For this analysis, 50,197 LiDAR observations were used for model-fitting.  

A Landsat 8 top of atmosphere (TOA) reflectance product was procured for the BCEF area for June of 2015. The June 2015 image was preferred to the June 2014 image due to excessive cloud cover in the 2014 image. A tasseled cap transformation was applied to the raw Landsat 8 TOA reflectance bands to obtain \emph{brightness}, \emph{greenness}, and \emph{wetness} tasseled cap indices \citep{hasan2014}. These tasseled cap indices are used as covariates in the subsequent analysis. 

Further details regarding the dataset and the ensuing analysis are provided in Section~\ref{sec:lidar}. 

\section{Modeling Strategy}\label{sec:model}

As stated before, our goal is to model and generate uncertainty equipped predictions of forest variables, making use of information contained in LiDAR signals. Consider a LiDAR signal, $\bz(\cdot)$, observed at a finite collection of locations $\oset_{z}=\lrb{\bs_{1},\ldots,\bs_{n_z}}$, and a set of forest outcomes, $\by(\cdot)$, is observed at locations in the set $\oset_{y}=\lrb{\br_1,\ldots,\br_{n_y}}\subset \oset_z$. Furthermore, let $\oset_{\emptyset}=\lrb{\bt_1,\ldots,\bt_{n_{\emptyset}}}$ denote a set of locations where neither LiDAR signals nor forest outcomes are available but where prediction is of interest.  Thus, the set of locations where both LiDAR and forest outcomes are to be mapped to corresponds to $\oset = \lrp{\oset_z \cup \oset_\emptyset}$, with $\oset \subset \mcal{D}\subset\mathbb{R}^2$, where $\mcal{D}$ is the spatial domain of interest.  Note that although we mention above that $\bz(\cdot)$ and $\by(\cdot)$ are ``observed'' at locations in $\oset_z$ and $\oset_y$, respectively, we allow for missing values that are to be imputed in these sets.  We make this distinction because locations where imputation is performed are part of the model fitting, whereas for locations in $\oset_\emptyset$ predictions are drawn \emph{ex post facto} from the posterior predictive distribution; more detail is provided in Section \ref{ssec:postpred}.

The LiDAR signals are high-dimensional vectors of measurements in $\mathbb{R}^{h_z}$, whereas the forest outcomes are relatively small-dimensional vectors (i.e., $h_y << h_z$), assumed to have support on $\mathbb{R}^{h_y}$. Forest outcomes and LiDAR signals are strongly dependent on each other; LiDAR signals vary with the composition of a forest, and, as a plethora of examples in the literature have demonstrated \citep{ene2018,finley2014,nelson2017} variability in forest outcome variables can be partially explained by LiDAR characteristics.

\subsection{Linking LiDAR and forest inventory data}\label{subsec:jointmod}

We seek to connect forest outcomes and LiDAR signals as a two-step process. First, we formulate a generative model to extract the spatial signature from the LiDAR data at locations in $\oset_z$, which can also be used to interpolate LiDAR signals in $\oset_\emptyset$. Along with other spatially referenced predictors, the LiDAR spatial signatures for locations in $\oset_y$ are used as predictors to build the model for forest outcomes. Moreover, a component that captures spatial variation exclusive to the forest outcomes can also be specified if required.   For $\bs\in\mcal{D}$ this two stage model is given by
\bea
\text{\sf{Stage 1: }}\bz(\bs) &=& \bX_z(\bs)'\bbeta_z + \bw^\star(\bs) + \bepsilon_z(\bs),\label{eq:inimodz}\\
\text{\sf{Stage 2: }}\by(\bs) &=& \bX_y(\bs)'\bbeta_y +  \mbs{\Upsilon}\bw^\star(\bs) + \bv^\star(\bs) + \bepsilon_y(\bs) \label{eq:inimody}
\eea
Note the influence of $\bz(\bs)$ over $\by(\bs)$ in \eqref{eq:inimody} is solely exerted through its spatial component $\bw^\star(\bs)$. There are several arguments in favor of this approach, as opposed to plugging in $\bz(\bs)$ or $\bmu_z(\bs)=\bX_z(\bs)'\bbeta_z + \bw^\star(\bs)$ directly as covariates into \eqref{eq:inimody}. Among them, and most importantly for our setting, $\bz(\bs)$, $\bmu_z(\bs)$ and $\bw^\star(\bs)$ are all high-dimensional objects, using $\bw^\star(\bs)$ facilitates reducing the dimensionality of the problem by casting it under the factor model structure, as shown in Section~\ref{subsec:comvia}. Additionally, the elements within $\bz(\bs)$ are strongly correlated and hence multicollinearity issues would arise if it was included directly in \eqref{eq:inimody}.

In \eqref{eq:inimodz} and \eqref{eq:inimody} the terms $\bX_z(\bs)^\prime\bbeta_z$ and $\bX_y(\bs)^\prime\bbeta_y$ capture large scale variation. For $\kappa \in\lrb{z,y}$, $\bX_{\kappa }(\bs)'$ represents a fixed $h_\kappa \times p_\kappa $ block-diagonal matrix of spatially referenced predictors, where $p_\kappa =\sum_{j=1}^{h_\kappa } p_{\kappa ,j}$, having as its $j$th diagonal block the length $p_{\kappa ,j}$ vector $\bx_j^{\kappa }(\bs)^\prime$. The length $p_\kappa$ vector $\bbeta_\kappa$  corresponds to the regression coefficients associated to $\bX_{\kappa}(\bs)'$. The vectors $\bw^\star(\bs)$ and $\bv^\star(\bs)$ are $h_z$ and $h_y$ dimensional zero-centered stochastic processes over $\mcal{D}$, respectively. The process $\bw^\star(\bs)$ captures the spatial variation of $\bz(\bs)$, while $\bv^\star(\bs)$ synthesizes additional spatial variation in the forest outcomes. The $h_y\times h_z$ matrix $\bUps$ connects the spatial information extracted from the LiDAR model into the forest outcomes model.  The vectors $\bepsilon_z(\bs)\sim \tN_{h_z}(\0,\bPsi_z)$ and $\bepsilon_y(\bs)\sim \tN_{h_y}(\0,\bPsi_y)$ represent uncorrelated random errors (i.e., $\bPsi_z$ and $\bPsi_y$ are diagonal) at finer scales.  

Implementing the modeling strategy above directly  is challenging due to the high-dimensionality of the LiDAR signals ($h_z\sim 50$) and the massive number of spatially dependent observations ($n\sim 10^5$), impossible to attempt with common computing resources.  In the following section, we formulate a viable alternative to model \eqref{eq:inimodz} and \eqref{eq:inimody}.

\subsection{The Spatial Factor NNGP Model}\label{subsec:comvia}

To make models \eqref{eq:inimodz} and \eqref{eq:inimody} tractable with limited computing power, we combine a dimension reduction approach and a sparsity inducing technique. In particular,  we introduce the spatial factor NNGP model (SF-NNGP), which brings together the spatial factor model (SFM) structure \citep{schmidt2003bayesian,finley2008bayesian,zhang2007maximum,Ren2013} with Nearest Neighbor Gaussian processes (NNGPs) \citep{Datta2016,Datta2016a,Datta2016c}. 

While the SFM structure enables the analysis of high-dimensional response vectors by using linear combinations of a relatively small number of independent stochastic processes, NNGPs make possible fitting spatial process models when the number of spatial observations is forbiddingly large.  NNGPs approximate the \emph{parent} (dense) GP by using the natural representation of sparsity provided by graphical models \citep{lauritzen96,murphy2012}, this by assuming conditional independence---where conditioning is on the nearest neighbors---with locations outside of the neighbor set.  The result is a proper (but sparse) GP that accurately approximates the original dense  GP.  In contrast to other sparsity inducing approaches, NNGPs allow for interpolation at unobserved locations, and can be used to make full inference on model parameters, including the latent processes.  Combining the SFM structure with NNGPs provides a methodology capable of coping simultaneously with high-dimensional response vectors and a large number of spatially dependent observations.  

Under the traditional SFM structure, the spatial dependence is introduced by defining the spatial process as $\bw^\star(\bs)=\bLambda\bw(\bs)\sim\text{GP}(\0,\mcal{H}(\cdot \given \bphi))$, where $\bLambda$ is a factor loadings matrix (commonly tall and skinny) and $\bw(\bs)$ is a small-dimensional vector of independent spatial GPs, providing the non-separable multivariate cross-covariance function given by 
\bea
\mcal{H}(\bh\given\bphi)&=&\text{cov}(\bLambda\,\bw(\bs), \bLambda\,\bw(\bs+\bh))\nonumber\\
&=& \sum_{k=1}^{q_w} \mcal{C}_k(\bh\given\phi_k) \blambda_k \blambda_k^\prime, \label{covmat}
\eea
for locations $\bs,\bs+\bh\in \mcal{D}$.  Here, $\mcal{C}_k(\bh|\phi_k)$'s are univariate parametric correlation functions, and $\blambda_k$ is the $k$th column of $\bLambda$. This cross-covariance matrix is induced by $q$-variate ($q\leq l$) spatial factors $\bw(\bs)$ with \emph{independent} components $w_k(\bs)\sim \text{GP}(0,\mcal{C}_k(\cdot\given \phi_k))$.

As such, models \eqref{eq:inimodz} and \eqref{eq:inimody} can be reformulated as SF-NNGPs by characterizing the spatial processes $\bw^\star(\bs)$ and $\bv^\star(\bs)$ as
\bea
\bw^\star(\bs) = \bLambda_z \bw(\bs)&\text{and}&\bv^\star(\bs) = \bGamma \,\bv(\bs), \label{eq:spfwv}
\eea
where the matrices $\bLambda_z=((\lambda_{hk}^{(z)}))_{h_z \times q_w}$, and $\bGamma=((\gamma_{lr} ))_{h_y\times q_v}$ correspond to the factor loadings matrices, and the new spatial factors for $\bs\in\mcal{D}$ are given by 
\bean
\bw(\bs) &\sim& \prod_{k=1}^{q_w} \nngpw{}{}{k}{w},\text{ and }\\
\bv(\bs) &\sim& \prod_{r=1}^{q_v}\nngpw{}{}{r}{v}.
\eean
The notation $\nngpw{}{}{k}{w}$ and $\nngpw{}{}{r}{v}$ denotes the Nearest Neighbor Gaussian Processes derived from the parent processes $\gpw{}{}{k}{w}$ and $\gpw{}{}{r}{v}$, respectively. Here, $\mcal{C}(\cdot\given \phi)$ represents the spatial correlation function with spatial decay parameter $\phi$. The factor model representation in \eqref{eq:spfwv} leads to a massive reduction in the dimensionality of the problem since the spatial factors $\bw(\bs)=(w_k(\bs): 1\leq k\leq q_w)$ and $\bv(\bs)=(v_{r}(\bs): 1\leq r\leq q_v)$, have dimensions $q_w << h_z$ and $q_v\leq h_y$. 

Bringing these elements together, and letting $\bLambda_y=\bUps \bLambda_z=((\lambda_{lk}^{(y)}))_{h_y \times q_w}$, a computationally viable version of \eqref{eq:inimodz} and \eqref{eq:inimody} is
\bea
\text{\sf{Stage 1: }}\bz(\bs) &=& \bX_z(\bs)'\bbeta_z + \bLambda_z \bw(\bs)+ \bepsilon_z(\bs) \label{eq:fnngpz}\\
\text{\sf{Stage 2: }}\by(\bs) &=& \bX_y(\bs)'\bbeta_y + \bLambda_y \bw(\bs) +  \bGamma\bv(\bs) + \bepsilon_y(\bs), \label{eq:fnngpy}
\eea

In general, additional constraints are required for factor models to be identifiable \citep{anderson2003}.  Identifiability for spatial factor models can be achieved either by making the upper triangle of the loadings matrix equal to 0 and its diagonal elements all equal to 1  \citep{Geweke1996,Lopes2004,Aguilar2010}, or as in \citet{Ren2013}, by fixing the sign of one element in each column of the factor loadings matrix, while enforcing an ordering constraint among the spatial decay parameters of the univariate correlation functions.  We choose to ensure rotation and scale identifiability by using the former approach.

With the SFM structure in place, introducing the NNGP reduces the expensive ($\sim$$n_z^3 q_w$ and $\sim$ $n_y^3 q_v$) calculation required to invert the dense covariance matrices from the parent GPs, by $n_z q_w$ and $n_y q_v$ parallel operations, each of order $m^3$.  Here, $m$ is the number of neighbors considered for the NNGP with $m<<n_y\leq n_z$. In simulations, \citet{Datta2016} found that in most cases $10 \leq m\leq 20$ provides an excellent approximation to the parent process; thus, the number of operations required is nearly linear in $n$.  

For completeness, additional details regarding SFMs and NNGPs, as well as the sampling algorithm, are included in the online supplement. For a more thorough treatment of SFM's we refer the reader to \citet{Ren2013,Genton2015}, and for NNGPs to \citet{Datta2016a}. 

\subsection{Prior Specification and Hierarchical Formulation}\label{ssec:priors}

Importantly, models \eqref{eq:fnngpz} and \eqref{eq:fnngpy} are fitted separately so the $\bw(\bs)$'s exclusively capture the spatial signal present in the LiDAR signals.  However, using plug-in estimates for $\bw(\bs)$ (e.g., the posterior means)  in \eqref{eq:fnngpy} disregards the uncertainty present in the LiDAR spatial signal.  Thus, to propagate this uncertainty through the forest outcome predictions, at each iteration of the from the Markov Chain Monte Carlo (MCMC) algorithm for $\by(\bs)$, we draw a sample for $\bw(\bs)$ ($\bs\in \oset_y$) MCMC samples obtained when fitting model \eqref{eq:fnngpz}. 

As mentioned in the previous section, the stochastic processes that capture the spatial structure are assumed to follow NNGPs.  Given that the NNGP is a proper Gaussian Process, at a finite collection of locations the NNGPs considered induce zero-centered multivariate normal priors with covariance matrices given by $\tbC^{(w)}$ and $\tbC^{(v)}$, respectively.  Additionally,  we use suitably noninformative priors for all other parameters, which make for a direct sampling strategy.  

In particular, we assume that $\bbeta$ is either flat or conjugate normal.  The matrices $\bGamma$ and $\bLambda_z$ are constrained as described above, with elements below the diagonal assumed to be standard normal. All elements in $\bLambda_y$ are also assumed to follow a standard normal distribution. The diagonal entries in $\bPsi_z$ and $\bPsi_y$ are assigned half-$t$ priors. Lastly, we assume uniform priors for the elements of the spatial decay vectors $\bphi_w=(\phi_{w,k}: 1\leq k \leq q_w)$  and $\bphi_v=(\phi_{v,r}: 1\leq r \leq q_v)$, in the interval $\lrp{-\log{0.05}/\zeta_{\text{max}}, -\log{0.01}/\zeta_{\text{min}}}$, where $\zeta_{\text{min}}$ and $\zeta_{\text{max}}$ are the minimum and maximum distances across all locations.  Given that $\bphi_z$ and $\bphi_y$ are not conjugate with their corresponding likelihood, these are sampled with random walk Metropolis steps.

The joint posterior densities for the first and second stages of the algorithm are proportional to
{\small
\bea
\textsf{Stage 1:}&&\hfill\nonumber\\
&&\pi(\bphi_w) \; \tN_{n_z q_w}(\bw_{\oset_z} | \0, \tbC^{(w)}) \;  \lrp{ \prod_{k=1}^{q_w} \prod_{j>k}^{h_z}  \tN(\lambda_{jk}^{(z)} | 0, 1)}  \nonumber \\
&& \qquad \times \; \pi(\bbeta_z) \lrp{\prod_{j=1}^{h_z} \mcal{IG}(\psi_j^z | \nu/2, \nu / a_{z,j}) \mcal{IG}(a_{z,j} | 1/2, 1 / A^2)} \nonumber\\
&&{}\qquad \times \lrp{\prod_{\bs_i\in\oset_z} \tN_{h_z}(\bz(\bs_i) | \bX_z(\bs_i)^\prime\bbeta_z + \bLambda_z \bw(\bs_i), \bPsi_z )},\label{eq:jointZ}
\eea
\bea
\textsf{Stage 2:}&&\hfill\nonumber\\
&&\pi(\bphi_v) \; \tN_{n_y q_v}(\bv_{\oset_y} | \0, \tbC^{(v)})\; \lrp{ \prod_{k=1}^{q_w} \prod_{j=1}^{h_y}  \tN(\lambda_{jk}^{(y)} | 0, 1)}  \; \lrp{ \prod_{r=1}^{q_v} \prod_{j>r}^{h_y}  \tN(\gamma_{jr} | 0, 1)} \nonumber \\
&& \qquad \times \; \pi(\bbeta_y) \lrp{\prod_{j=1}^{h_y} \mcal{IG}(\psi_j^y | \nu/2, \nu / a_{y,j}) \mcal{IG}(a_{y,j} | 1/2, 1 / A^2)} \nonumber\\
&&{}\qquad \times \lrp{\prod_{\bs_i\in\oset_y} \tN_{h_y}(\by(\bs_i) | \bX_y(\bs_i)^\prime\bbeta_y + \bLambda_y \bw(\bs_i) + \bGamma \bv(\bs_i), \bPsi_y )},\label{eq:jointY}
\eea
}
where, the vectors $\bw_{\oset_z} = \lrp{\bw(\bs_i)^\prime: \bs_i\in\oset_z}^\prime$, and $\bv_\oset = \lrp{\bv(\bs_i)^\prime: \bs_i\in\oset_y}^\prime$, such that
\bea
\tN_{n_z q_w}(\bw_\oset | \0, \tbC^{(w)}) &=& \prod_{\bs_i\in\oset_z} \tN_{q_w}(\bw(\bs_i) \given \bB_{i}^{(w)} \bw_{N(i)}, \bF_{i}^{(w)}),\text{ and}\nonumber \\
\tN_{n_y q_v}(\bv_\oset | \0, \tbC^{(v)}) &=& \prod_{\bs_i\in\oset_y} \tN_{q_v}(\bv(\bs_i) \given \bB_{i}^{(v)} \bv_{N(i)}, \bF_{i}^{(v)}).\label{priorsNNGP}
\eea

The expressions on the right hand side of \eqref{priorsNNGP} result from the construction of the NNGP (see online supplement). For an $m$-neighbor NNGP, denote by $m_i=\text{min}\lrb{m,i-1}$ the number of neighbors for location $\bs_i$.  The index set $N(i)$ for location $\bs_i\in\oset_z$ contains its $m_i$ nearest neighbors; thus $\bw_{N(i)}$ corresponds to the vector $\lrp{\bw(\bs_j)': \bs_j\in N(i)\subset \oset_z}'$.  The neighbor sets are defined analogously for the $\bv(\bs_i)$'s.  Letting $u \in \lrb{w, v}$,  $\bB_i^{(u)}$ denotes the $q_{u}\times m_i q_{u}$ block matrix, with $q_{u}\times q_{u}$ diagonal blocks containing the kriging weights for the $q_u$ spatial factors for each neighbor. Also, $\bF_i^{(u)}$ corresponds to the $q_u \times q_u$ diagonal matrix with the variances for the $q_u$ spatial factors conditioned on the neighbor set $N(i)$ (see Section \ref{subsec:nngp} in the supplement for details regarding $\bB_i^{(u)}$ and $\bF_i^{(u)}$).  Lastly, the parameters $\lrb{a_{y,j}}_{j=1}^{h_y}$ and $\lrb{a_{z,k}}_{k=1}^{h_z}$ complete the hierarchical representation of the half-$t$ prior distribution assumed for $\psi_j^y$ and $\psi_k^z$, respectively, and the hyperparameter $A$ is simply chosen to be some large value (say, 100). 

Due to prior conjugacy, the full conditional densities for all parameters, except for those of $\bphi_w$ and $\bphi_v$, can be sampled using simple Gibbs steps.  Further details on the sampling algorithm are deferred to the online supplement.

\subsection{Imputation and Prediction}\label{ssec:postpred}

As mentioned before, LiDAR signals are collected over the large spatial region $\oset_z$, whereas forest outcome observations are confined to the smaller subset of locations in $\oset_y$.  Additionally, there are relevant out-of-sample locations where neither LiDAR nor forest outcomes are observed, $\oset_\emptyset$. And finally, there are some locations within the corresponding reference sets $\oset_z$ and $\oset_y$ that have some or all missing outcomes. It is thus essential for this modeling effort to provide the means to accurately impute the missing values in $\oset_z$ or $\oset_y$, and generate LiDAR predictions in $\oset_\emptyset$ and forest outcome predictions within $\oset_\emptyset \cup (\oset_z\setminus\oset_y)$. Given the NNGP formulation, both imputation and out-of-sample prediction are remarkably inexpensive.

Imputation is straightforward. Let $\bs_{\bullet}\in\oset_z$ be a location where $\bz(\bs_{\bullet})$ is missing. Then $\bz(\bs_{\bullet})$ is drawn as part of the sampling algorithm from $\tN_{h_z}( \bX_z(\bs_\bullet)^\prime\bbeta_z + \bLambda_z \bw(\bs_\bullet), \bPsi_z )$, where $\bw(\bs_\bullet)$ is sampled from the full conditional posterior density in Equation \eqref{eq:postWun} of the online supplement. For a missing value $\by(\bs_\bullet)$, where $\bs_\bullet\in \oset_y$, the procedure is analogous using the full conditional posterior for $\bv(\bs_\bullet)$ and the likelihood for $\by(\bs_\bullet)$.

The procedure to predict a new LiDAR observation $\bz(\bs_\circ)$, $\bs_\circ \in \oset_\emptyset$, begins by sampling the spatial factor $\bw(\bs_\circ)$ from $\tN_{q_w}(\bB_\circ^{(w)} \bw_{N(\bs_\circ)}, \bF^{(w)}_{\circ})$, with $\bB_\circ^{(w)}$ and $\bF^{(w)}_{\circ}$ defined as before. Note that the nearest neighbor set $N(\bs_\circ)$ is assumed to be in $\oset_z$.  Then, we draw $\bz(\bs_\circ)\given \bz_{\oset_z}$ from $\tN_{h_z}( \bX_z(\bs_\circ)^\prime\bbeta_z + \bLambda_z \bw(\bs_\circ), \bPsi_z )$.  This is done by conditioning on the posterior samples of $\lrb{\bbeta_z,\bLambda_z,\bPsi_z,\bphi_w}$ obtained from the fitting algorithm. 

To predict the forest outcomes $\by(\bs_\circ)$ at $\bs_\circ\in \oset_\emptyset \cup (\oset_z\setminus\oset_y)$, first we generate samples of $\bv(\bs_\circ)\sim \tN_{q_v}(\bB_\circ^{(v)} \bv_{N(\bs_\circ)}, \bF^{(v)}_{\circ})$. Given that $\by(\bs_\circ)$ depends on $\bw(\bs_\circ)$, we combine the posterior draws of $\lrb{\bbeta_y,\bLambda_y,\bGamma,\bPsi_y,\bphi_v}$ with those of $\bw(\bs_\circ)$, obtained when predicting $\bz(\bs_\circ)$, and draw predicted values for $\by(\bs_\circ)\given \by_{\oset_y}$ from $\tN_{h_y}( \bX_y(\bs_\circ)^\prime\bbeta_y + \bLambda_y \bw(\bs_\circ) + \bGamma \bv(\bs_\circ), \bPsi_y )$.

\section{Simulation: Recovering Low-dimensional Structure}\label{sec:sims}

In the following simulation exercise we focus exclusively on the high-dimensional component (i.e., the first stage) of the model described above. 
The simulation below was devised to illustrate the ability of our approach to recover true low-dimensional structure when data is generated from a low-dimensional SFM with dense spatial factors. 

We generate a synthetic dataset for $h_z=50$ outcomes in $n_z=10,000$ locations from the spatial factor model
$\bz(\bs)=\bX_z(\bs)'\tilde{\bbeta}_z + \tilde{\bLambda}_z \tilde{\bw}(\bs)+\tilde{\bepsilon}_z(\bs).$
Here, $\bX_z(\bs)'$ is a $50\times 150$ block-diagonal matrix of predictors, and $\tilde{\bbeta}_z$ is the vector of regression coefficients, both defined as before. We consider the same three predictors for all outcomes. The spatial factors $\tilde{\bw}(\bs) \sim\prod_{k=1}^8\text{GP}(0,\mcal{C}(\cdot\given \tilde{\phi}_{k}^{z}),$ where $\mcal{C}(\cdot\given\tilde{\phi}_k^z)$ is an exponential correlation function with decay parameter $\tilde{\phi}_k^z$.  Additionally, for identifiability we assume that the $50\times 8$ factor loadings matrix $\tilde{\bLambda}_z$ has zeros in the upper triangle and ones along the diagonal. Finally, $\tilde{\bepsilon}_z\sim \tN_{h_z}\lrp{\0,\tilde{\bPsi}_z}$, with $\tilde{\bPsi}_z=\text{diag}(\tilde{\psi}_k^z: k=1,\ldots,8)$.

We assess the ability of model \eqref{eq:fnngpz} to recover model parameters from the true data generating process, impute missing outcomes, and predict at out-of-sample locations. The SF-NNGP model was fitted for $q_w\in\lrb{3, 5, 8, 10}$ spatial factors and assuming $m=10$ neighbors.  Out of the $10,000$ locations, we assume all 50 outcomes to be missing in 200 locations chosen at random, and impute them. Additionally we hold out $n_0=500$ locations for out-of-sample prediction and model validation.

The first result worth highlighting is the gains in computational efficiency provided by the SF-NNGP.  For this particular simulation exercise---a relatively  computationally challenging problem---fitting the largest model considered (i.e., $q_w=10$) with 50,000 MCMC iterations, on a Linux server with Intel i7 processor (two 8-core) and 16 GB of memory, the runtime was 4.88 hours. As shown below, the proposed approach is able to recover the true model parameters, accurately impute missing data and generate precise predictions, all of these equipped with suitable uncertainty estimates.

For all values of $q_w$ the SF-NNGP accurately recovered the regression coefficients $\tilde{\bbeta}_z$ for all predictors and responses (Figure \ref{fig:sim1betas1} in the online supplement).   In contrast, the quality of the estimates for the small-scale variance components $\tilde{\psi}_k^z$'s was compromised when $q_w$ was lower than the true number of spatial factors. This behavior is expected, for lower values of $q_w$ the $\psi_k^z$'s attempt to compensate for the additional signal that the spatial component with too few spatial factors is unable to capture (Figure \ref{fig:sim1taus} in the supplementary material). For $q_w=8$ and $q_w=10$, the coverage for $\tilde{\bpsi_z}$ was 88\% and 84\%, respectively, with all $\psi_k^z$ close to $\tilde{\psi}_k^z$ with tight 95\% credible sets.  

When $q_w\not=8$, the dimensions of the fitted $\bLambda_z$, $\bphi_w$, and $\bw(\bs)$ do not match those of their analogs in the true model.  Therefore, to assess the quality of fit for the spatial signal for all values of $q_w$ considered, we instead compare the fitted spatial component $\bw^\star(\bs)= \bLambda_z\bw(\bs)$, for $\bs\in\oset_z$, to that of the true model, given by $\tilde{\bw}^\star(\bs)= \tilde{\bLambda}_z\tilde{\bw}(\bs)$. 

\begin{figure}[H]
\centering
\subfigure[$q_w=3$]{\includegraphics[scale=0.35,trim={0 0 2.1cm 0cm},clip]{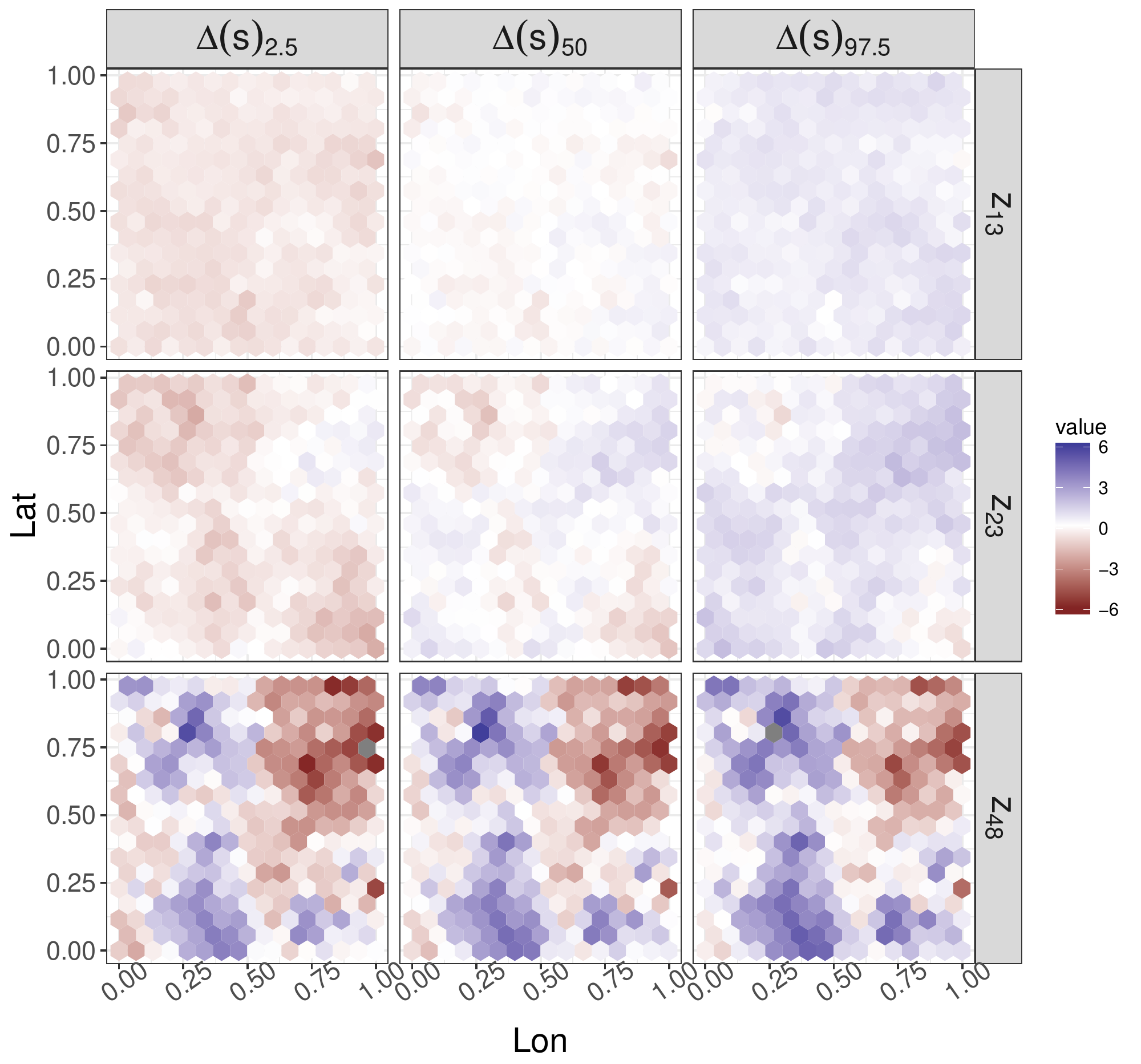}}
\subfigure[$q_w=5$]{\includegraphics[scale=0.35,trim={1.75cm 0 0 0cm},clip]{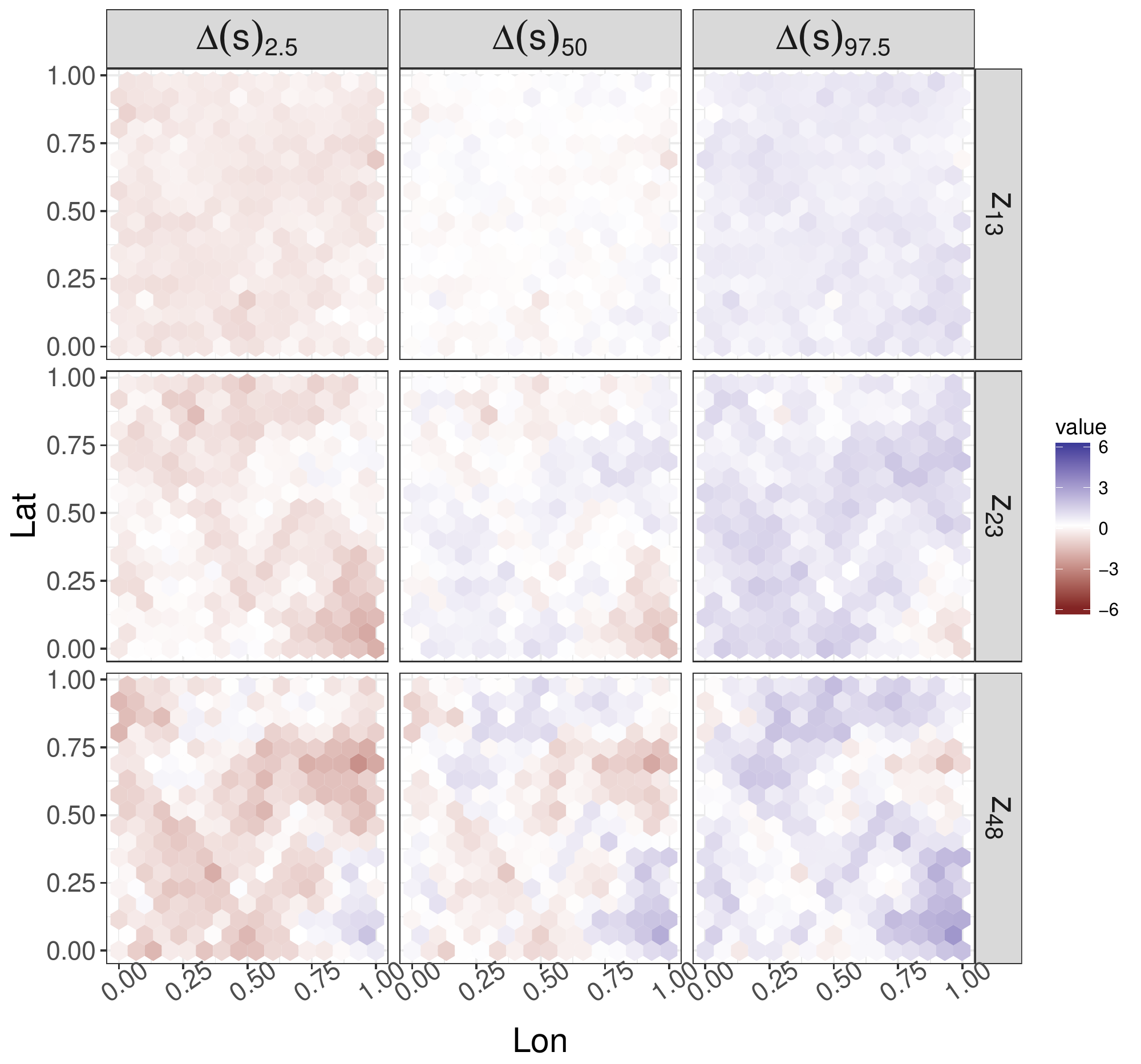}}
\subfigure[$q_w=8$]{\includegraphics[scale=0.35,trim={0 0 2.05cm 0cm},clip]{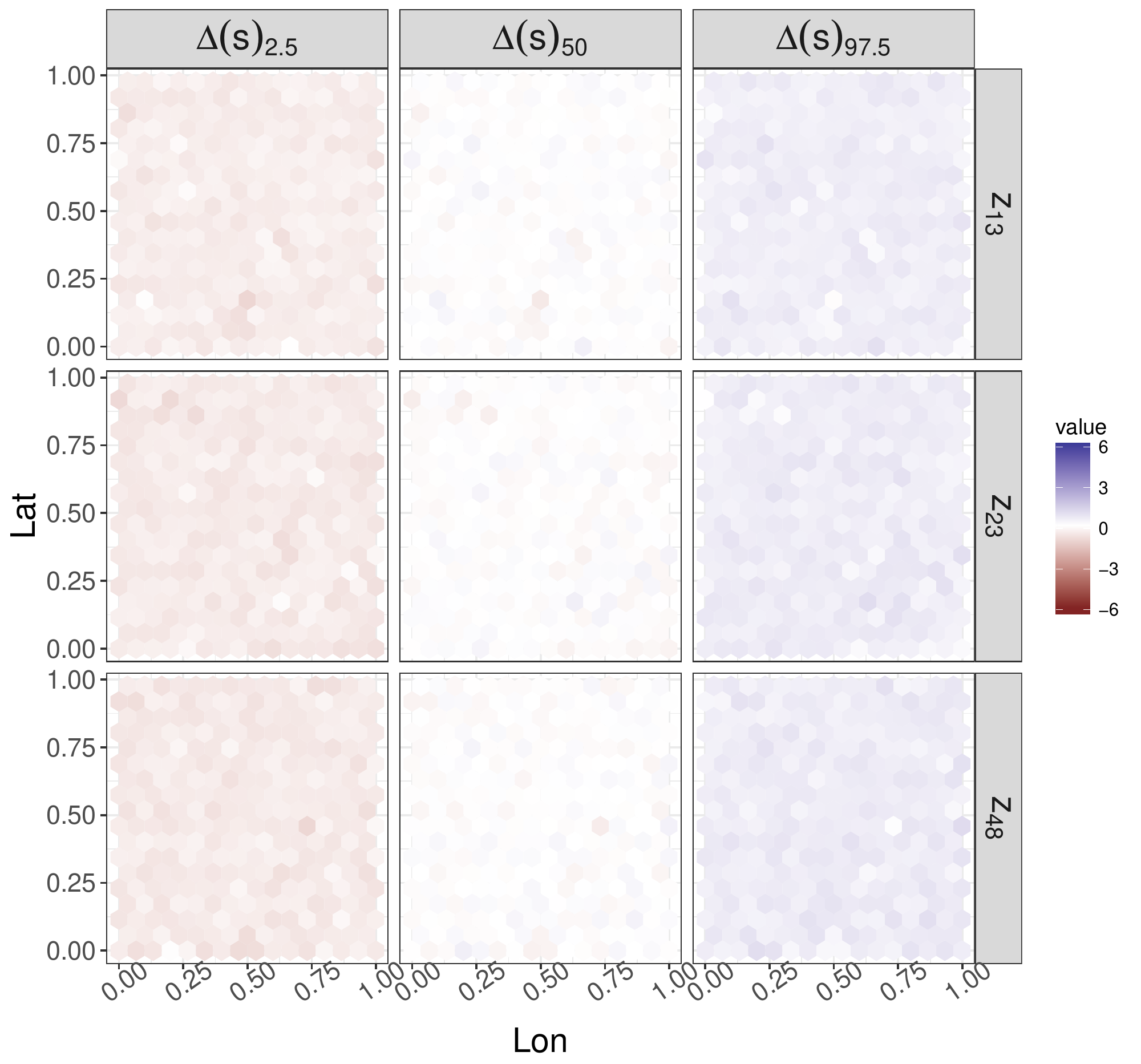}}
\subfigure[$q_w=10$]{\includegraphics[scale=0.35,trim={1.75cm 0 0 0cm},clip]{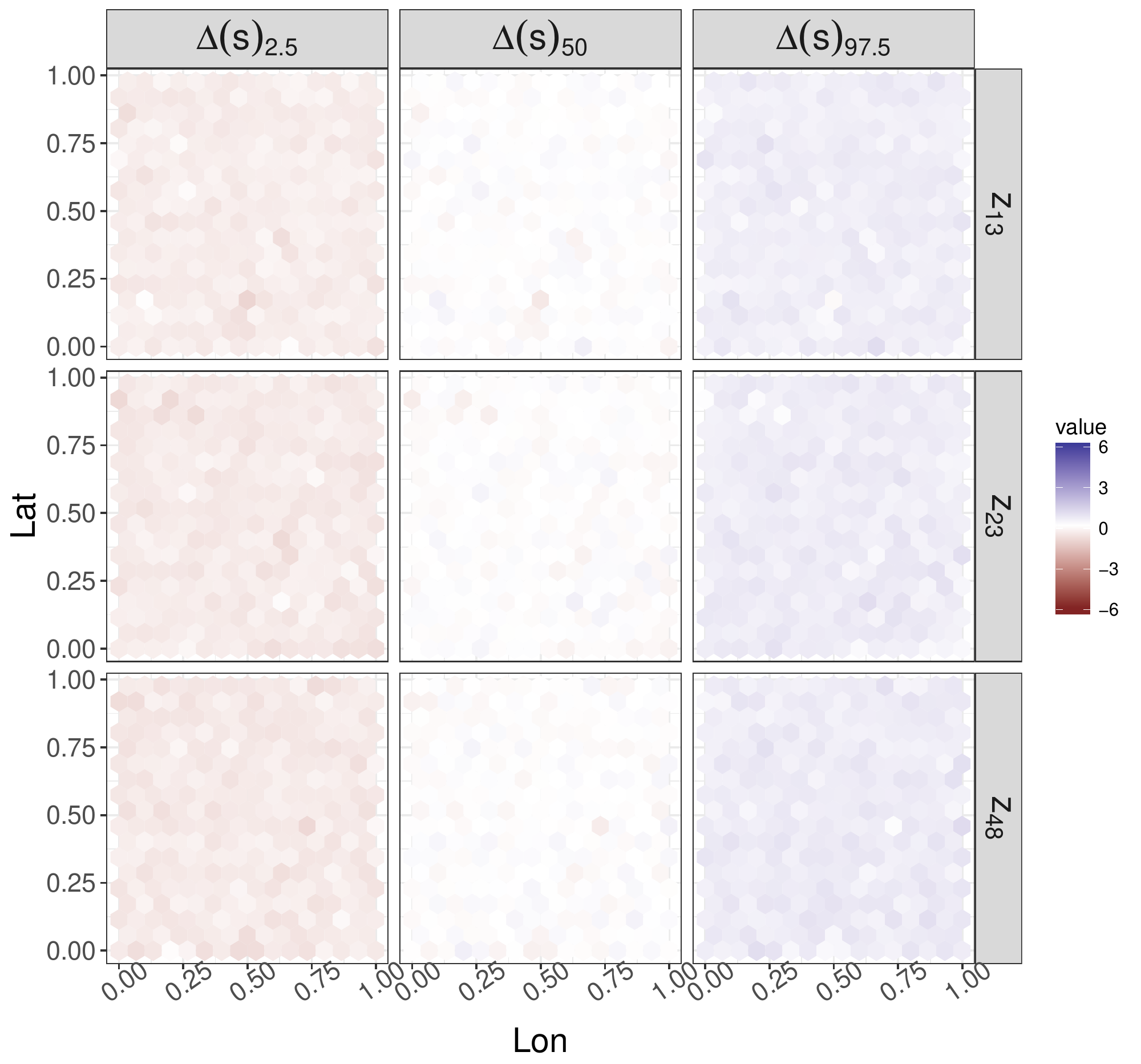}}
\caption{Fitted minus true spatial signal, $\Delta(\bs)=\bw^\star(\bs)-\tilde{\bw}^\star(\bs)$, for locations $\bs_{13}, \bs_{23}, \bs_{48}$.  From left to right the columns in each panel show percentiles 2.5, 50 and 97.5 for $\Delta(\bs)$, respectively. }\label{fig:SpCompq}
\end{figure}

For all locations in $\oset_z$ we calculate $\Delta(\bs)=\bw^\star(\bs)-\tilde{\bw}^\star(\bs)$ (fitted minus true spatial signal) for each MCMC draw of the parameters. For all $\bs\in\oset_z$ we obtained the median and 95\% credible set for $\Delta(\bs)$. To facilitate visualization, in Figure \ref{fig:SpCompq} we show the results for only  three responses selected at random from the 50 considered. The columns of each panel map quantiles 2.5, 50 and 97.5 for $\Delta(\bs)$ with 3 locations (13, 23 and 48) plotted by row. 
The fitted spatial signal when $q_w\in\lrb{3,5}$ recovers only partially the true signal, with coverages of 26.13\% and 42.06\%, respectively for $q_w=3$ and $q_w=5$.  When $q_w\in\lrb{8,10}$ the recovery of the spatial signal is astonishingly accurate, having over all responses 94.78\% coverage with $q_w=8$, and 94.18\% coverage with $q_w=10$.  

\begin{figure}[H]
\centering
\includegraphics[scale=0.47,trim={0.1cm 0 0.3cm 0.7cm},clip]{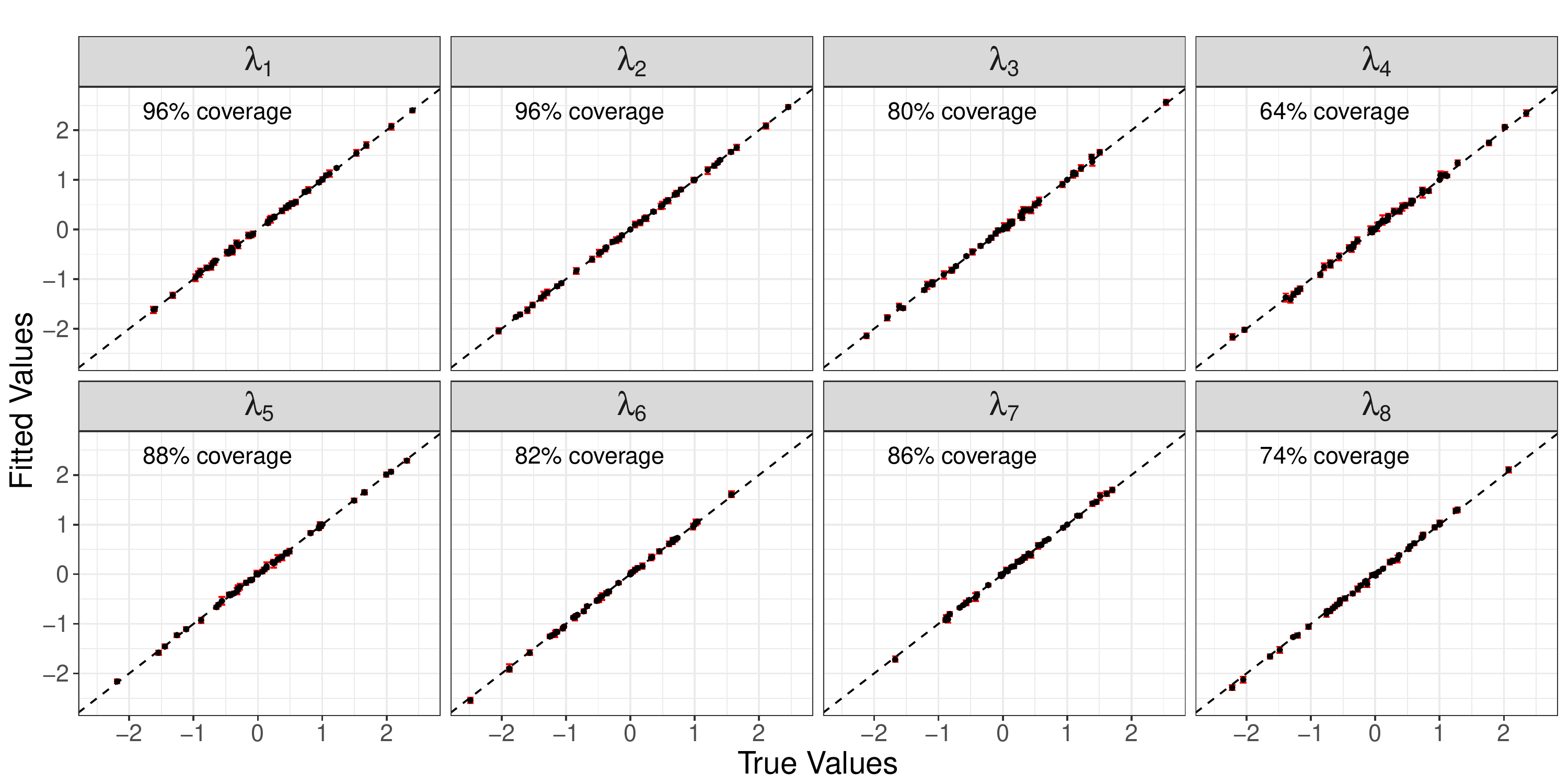}
\caption{Fitted vs true factor loadings matrix parameters (95\% credible sets and medians) for $q_w=8$.}\label{fig:simslambdaq8}
\end{figure}

\begin{figure}[H]
\centering
\includegraphics[scale=0.5,trim={0cm 0 0.3cm 0.7cm},clip]{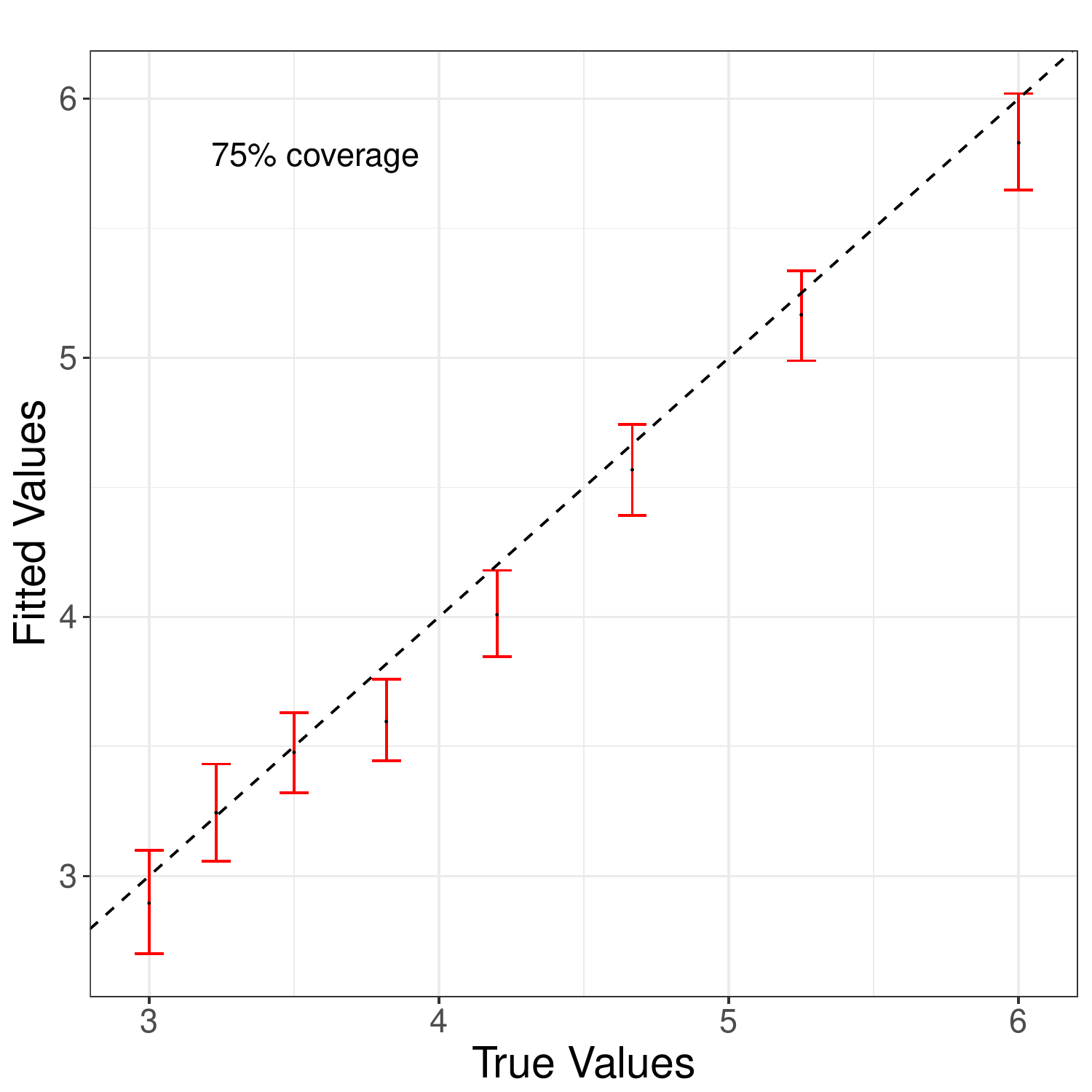}
\caption{Fitted vs true spatial decay parameters parameters (95\% credible sets and medians) for $q_w=8$.}\label{fig:simsphiq8}
\end{figure}

In addition to the previous results, it is also encouraging to find that when the dimension of the SF-NNGP model matches that of the true model, both the factor loadings ($\tilde{\bLambda}_z$) and the spatial decay parameters ($\tilde{\bphi}_z$) from the true spatial process can be recovered accurately (Figures \ref{fig:simslambdaq8} and \ref{fig:simsphiq8}).  

Model performance in terms of the accuracy of both imputation and prediction improves drastically as the number of factors approaches the truth -- see Figures \ref{fig:sim1impute} and \ref{fig:sim1pred} in the online supplement. 

Table \ref{tab:gof} provides a comparison as $q_w$ varies in the SF-NNGP using different measures of out-of-sample predictive performance. In particular, the continuous rank probability score (CRPS) \citep[Equation (21) in][]{Gneiting2007} and the root mean squared prediction error (RMSPE) \citep{yeniay2002} favor the model with $q_w=8$. The coverage of the 95\% credible intervals of the predictions was close to the nominal value for all $q_w$; however, the width of the interval rapidly decreases as $q_w$ approaches the true number of spatial factors. 

\begin{table}[ht]\caption{Out-of-sample prediction comparison across models with different number of spatial factors.}\label{tab:gof}
\bigskip
\centering
\begin{tabular}{rcccc}
  \toprule
$q_w$ & CRSP & RMSPE &  95\% Coverage & 95\% CI Width \\ 
  \midrule
3 & 0.85 & 1.61 & 95.82 & 6.14 \\ 
5 & 0.67 & 1.28 & 95.43 & 4.79 \\ 
8 & 0.45 & 0.83 & 94.78 & 3.10 \\ 
10 & 0.45 & 0.83 & 94.84 & 3.10 \\ 
   \bottomrule
\end{tabular}
\end{table}

Both the fitted values for the spatial signals and the out-of-sample predictions with $q_w=8$ and $q_w=10$ are practically indistinguishable from each other. Furthermore, the model with $q_w=8$ accurately recovers all the true factor loadings (Figure \ref{fig:simslambdaq8}).  Interestingly, with $q_w=10$, visual inspection of the estimates for columns 1 through 6 in $\bLambda_z$ indicate that this model accurately estimates the corresponding true parameter values (see Figure \ref{fig:sim1load2q10} in the online supplement).  However,  in this same model the estimated parameter values in columns 7 and 8 of $\bLambda_z$ display departures from their true values; and the 95\% credible sets for all the unconstrained elements in the 9th and 10th columns of $\bLambda_z$ contain zero (see Figure \ref{fig:sim1loadq10} in the online supplement).  These results provide guidance regarding the selection of the number of factors $q_w$ to use. As there is no gain in using the model with $q_w=10$ over the one with $q_w=8$ in terms of predictive accuracy or parameter fit, the results favor the more parsimonious model of the two.

\section{Modeling LiDAR Signals and Forest Structure}\label{sec:lidar}

Our focus in the subsequent analysis is to assess and interpret the utility of SF-NNGP spatial factors to explain variability in the three forest outcomes defined in Section~\ref{sec:data}, measured on the BCEF. Following the two stage model developed in Section~\ref{subsec:comvia}, we fit \eqref{eq:fnngpz} using $q_w\in\lrb{1, 2, 3, 4, 5, 6, 7, 8}$ spatial factors and $m=10$ neighbors to the BCEF LiDAR data comprising $n_z$=50,197 signals each of length $h_z$=57. The model mean included only an intercept. Prior specification followed Section~\ref{ssec:priors}, with the support for elements in $\bphi_w$ adjusted to match the BCEF spatial extent.

The $n_y$=197 locations with $h_y$=3 forest outcomes were used in the second stage model \eqref{eq:fnngpy}. To more clearly interpret the spatial factors' ability to explain variability in forest outcomes, we decided to avoid potential issues with spatial confounding \citep{hanks2015} and set $\bv(\bs)$ to zero. In practice, however, if our main objective is to maximize predictive performance then this residual spatial random effect should likely be included in the model. In addition to the spatial factors, the second stage model was informed by the three Landsat 8 tasseled cap predictor variables defined in Section~\ref{sec:data} which, along with an intercept, were included in $\bX_y(\bs)$. Importantly, these predictor variables are available across the entire BCEF, hence, given predicted values of the spatial factors at unobserved locations, we can create complete-coverage forest outcome maps.  

Posterior inference for all candidate models was based on three chains of 50,000 post burn-in MCMC samples. Chains converged by 20,000 MCMC iterations. Using the same computer configuration detailed in Section~\ref{sec:sims}, total runtime for the most demanding model, i.e., $q_w=8$, was $\sim$36 hours.

The eight candidate models, specified by $q_w$, were assessed based on their ability to inform forest outcome prediction. This was done by fitting each of the first stage models, then fitting their corresponding second stage models using data from 99 of the197 available locations in $\oset_{y}$. The three forest outcomes were then predicted for the remaining 98 out-of-sample locations. Scoring rules and other summaries of the posterior predictive distributions for the 98 out-of-sample locations are presented in Table~\ref{tab:xval-real}.

Increasing the number of spatial factors improves CRPS and RMSPE for each forest outcome Table~\ref{tab:xval-real}. Exploratory analysis showed gains in predictive performance were negligible beyond $q_w=4$ for AGB and $q_w=5$ for TD and BA. Given that the $q_w$=5 model generally yielded the ``best'' predictions, it was selected for exposition below.

\begin{table}[H]\caption{Cross-validation prediction summary for forest outcomes given increasing number of spatial factors $q_w$. Bold values identify lowest CRPS and RMSPE.}\label{tab:xval-real}
\bigskip
\centering
\small{
\begin{tabular}{cccccc}
  \toprule
 & $q_w$ & CRSP & RMSPE &  95\% Coverage & 95\% CI Width \\ 
  \midrule
  \multirow{4}{*}{AGB}
  &1&26.21& 51.37& 91.88 &161.24\\
 &2	&26.36& 52.02& 92.39& 162.14\\
 &3	&23.64& 46.95& 95.94& 155.71\\
 &4	&\textbf{23.53} &\textbf{46.93}& 93.91& 155.66\\
  &5	&24& 47.54& 96.45& 157.75\\
  &6	&24.47& 47.8& 94.92& 172.64\\
  &7	&24.75& 47.84& 95.43& 174.44\\
  &8	&24.76& 48.02& 96.45& 182.12\\
     \midrule
     \multirow{4}{*}{TD}
     	&1 &1017.7& 1980.62& 92.39& 6010.6\\
	&2 &1006.02& 1957.54& 93.4& 5944.81\\
	&3 &1007.72& 1954.87& 93.4& 6068.29\\
	&4 &997.32& 1955.2& 93.4& 6040.06\\
     &5 &\textbf{989.31}& \textbf{1930.76}& 94.92& 6182.2\\
     &6 &998.3 &1944.22& 94.42& 6223.73\\
     &7&	1005.26 &1965.81& 95.43& 6450.5\\
&8&	1004.36& 1955.08& 96.95& 6503.17\\
   \midrule
   \multirow{4}{*}{BA}
   	&1 &5.53& 10.29& 91.88& 36.34\\ 
	&2 &5.4& 10.01& 94.42& 36.85\\ 
	&3 &5.13& 9.54& 93.91& 35.16\\
	&4 &5.17& 9.62& 93.4& 36.21\\
   &5 &\textbf{5.16}& \textbf{9.58} &93.4& 36.51\\
   &6 &5.2 &9.59& 96.45 &38.62\\
   &7	&5.24 &9.73 &95.43 &38.34\\
&8	&5.27 &9.72 &94.42 &37.93\\
   \bottomrule
\end{tabular}
}
\end{table}





\begin{table}[H]\caption{Elements of $\bLambda_y$ median and 95\% credible intervals for the $q_w=5$ model. Bold entries indicate where the 95\% credible interval excludes zero.}\label{tab:lambda-real}
\bigskip
\centering
\begin{tabular}{cc}
  \toprule
  Parameter&50\% (2.5\%, 97.5\%) \\ 
  \midrule
 $\lambda_{\textsf{AGB},1}^{(y)}$  &\textbf{-6.65 (-8.89, -4.23)}\\
 $\lambda_{\textsf{AGB},2}^{(y)}$  &27.20 (-14.11, 65.14)\\
 $\lambda_{\textsf{AGB},3}^{(y)}$  &\textbf{-278.29 (-324.52, -232.28)}\\
  $\lambda_{\textsf{AGB},4}^{(y)}$  &-46.15 (-162.56, 75.91)\\
   $\lambda_{\textsf{AGB},5}^{(y)}$  &\textbf{-308.81 (-524.42, -90.45)}\\
     \midrule
 $\lambda_{\textsf{TD},1}^{(y)}$  &-1.77 (-21.35, 17.60)\\
 $\lambda_{\textsf{TD},2}^{(y)}$  &\textbf{-357.49 (-718.82, -7.86)}\\
     $\lambda_{\textsf{TD},3}^{(y)}$  &269.03 (-137.51, 667.62)\\
    $\lambda_{\textsf{TD},4}^{(y)}$  &\textbf{-1777.21 (-2696.67, -708.08)}\\
   $\lambda_{\textsf{TD},5}^{(y)}$  &\textbf{2457.52 (681.18, 4337.97)}\\
      \midrule
 $\lambda_{\textsf{BA},1}^{(y)}$  &\textbf{-2.93 (-3.94, -1.75)}\\
 $\lambda_{\textsf{BA},2}^{(y)}$  &-2.07 (-19.02, 15.79)\\
 $\lambda_{\textsf{BA},3}^{(y)}$  &\textbf{-98.64 (-119.79, -76.24)}\\
 $\lambda_{\textsf{BA},4}^{(y)}$  &\textbf{-72.00 (-120.60, -23.00)}\\
 $\lambda_{\textsf{BA},5}^{(y)}$  &-80.55 (-177.44, 20.51)\\
   \bottomrule
\end{tabular}
\end{table}

\begin{figure}[H]
\centering 
\includegraphics[scale=0.75,trim={0cm 0cm 0cm 0cm},clip]{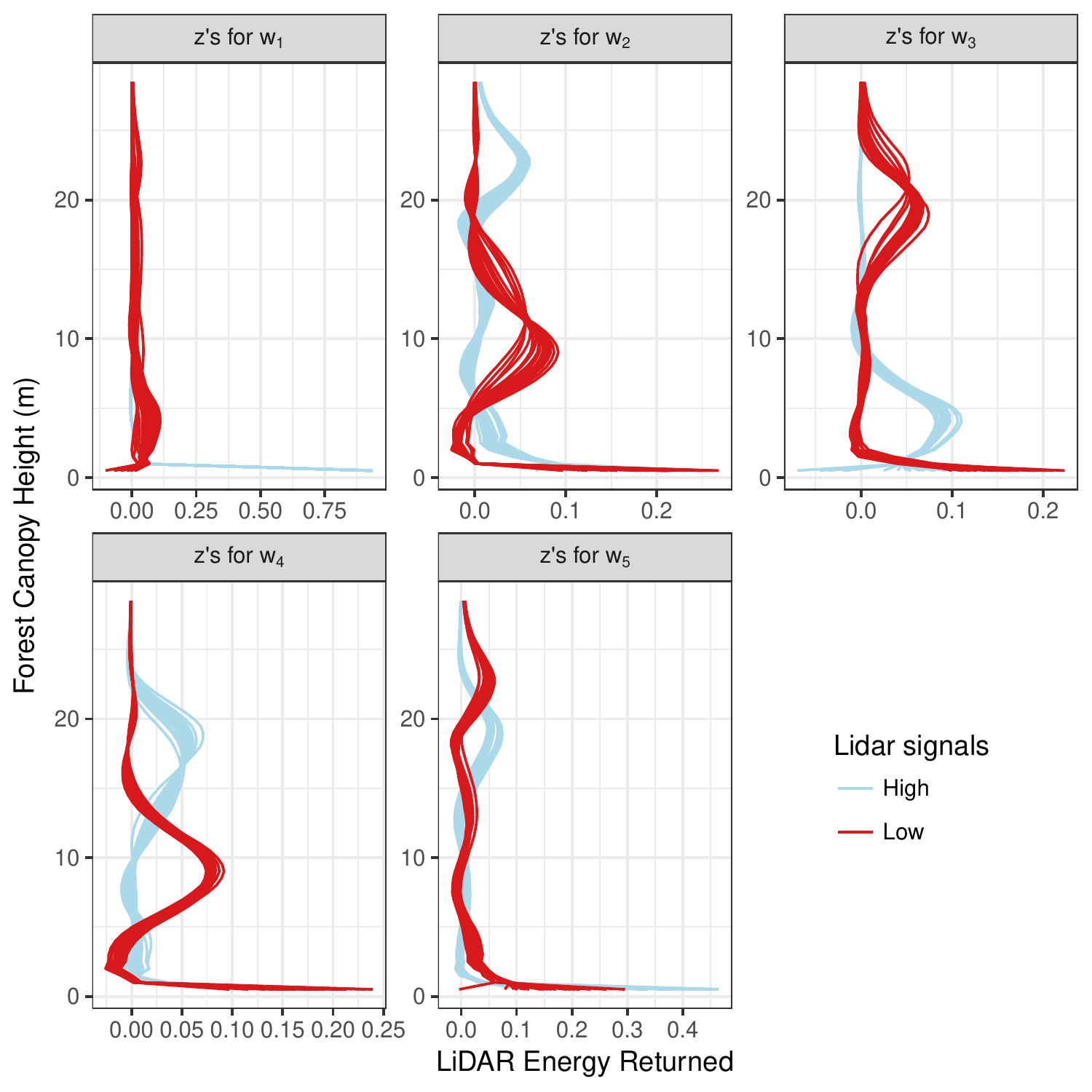}
\caption{Observed LiDAR signals with the 25 largest (\emph{High} in the legend) and 25 smallest (\emph{Low} in the legend) values of $\bw(\bs)$'s elements from the $q_w=5$ model.}\label{fig:signals-map}
\end{figure}

Table~\ref{tab:lambda-real} provides estimates for the second stage model's spatial factor regression coefficients, i.e., elements in $\bLambda_y$. These results show that several of the spatial factors explain a substantial portion of variability in the forest outcomes. It is, however, difficult to interpret the $\lambda^{(y)}$'s without a sense of what characteristic of $\bz(\bs)$ the spatial factors are capturing. When considered with estimates in Table~\ref{tab:lambda-real}, Figure~\ref{fig:signals-map} provides some biological interpretation of the spatial factors. Specifically, each panel in Figure~\ref{fig:signals-map} represents a spatial factor. The 50 lines in each panel are observed LiDAR signals with color corresponding to the 25 largest (blue lines) and 25 smallest (red lines) estimated spatial factor value.

There are some general biological relationships between forest canopy structure and AGB, TD, and BA. Very low maximum canopy height is indicative of a young regenerating forest (e.g., regrowth after a fire) that would be characterized by low AGB, high TD, low BA. If the majority of trees in a forest have a high canopy height then we expect high AGB, low TD, and high BA (i.e., few large diameter mature trees dominate the area). When the forest is characterized by trees of many different heights (i.e., tree crowns in several vertical strata) then we might expect moderate/high AGB, moderate TD, and moderate/high BA. Some of these expected relationships are observed when comparing Table~\ref{tab:lambda-real} and Figure~\ref{fig:signals-map}. For example, the top left panel in Figure~\ref{fig:signals-map} differentiates between regenerating forests and all other forest structure, i.e., blue lines show a spike of energy returned at or near ground level versus red lines which show the majority of the energy is returned at or above several meters. Hence negative regression coefficients $\lambda^{(y)}_{AGB,1}$ and $\lambda^{(y)}_{BA,1}$ in Table~\ref{tab:lambda-real}. The LiDAR signals shown in the top right panel in Figure~\ref{fig:signals-map} differentiates between young and old single cohort forests (i.e., all trees were regenerated around the same time and there is little vertical variation in canopy height); hence, negative  $\lambda^{(y)}_{AGB,3}$ and $\lambda^{(y)}_{BA,3}$ in Table~\ref{tab:lambda-real}. The top middle and bottom left panels in Figure~\ref{fig:signals-map} generally separate blue signal mature 20+ and $\sim$20 meter canopy height, respectively, from lower stature $\sim$10 meter canopy height forest. Consistent with the biological expectation, the negative $\lambda^{(y)}_{TD,2}$ and $\lambda^{(y)}_{TD,4}$ suggest forests associated with red LiDAR signals have higher tree density relative to the older taller forests.

\begin{figure}[H]
\centering 
\subfigure[AGB (Mg ha$^{-1}$)]{\includegraphics[scale=0.45,trim={0cm 3.0cm 0.3cm 3cm},clip]{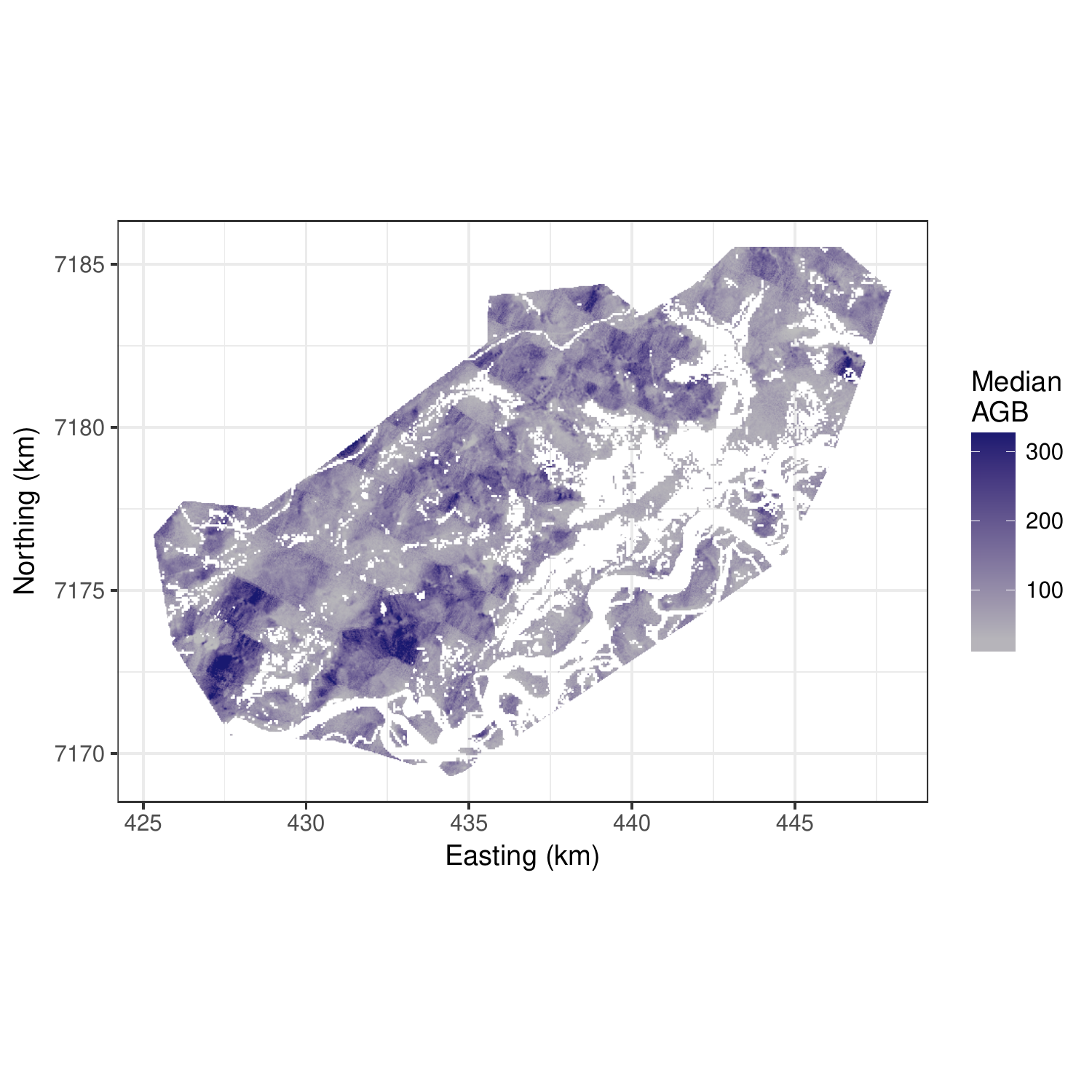}}
\subfigure[AGB 95\% CI Width (Mg ha$^{-1}$)]{\includegraphics[scale=0.45,trim={0cm 3.0cm 0.3cm 3cm},clip]{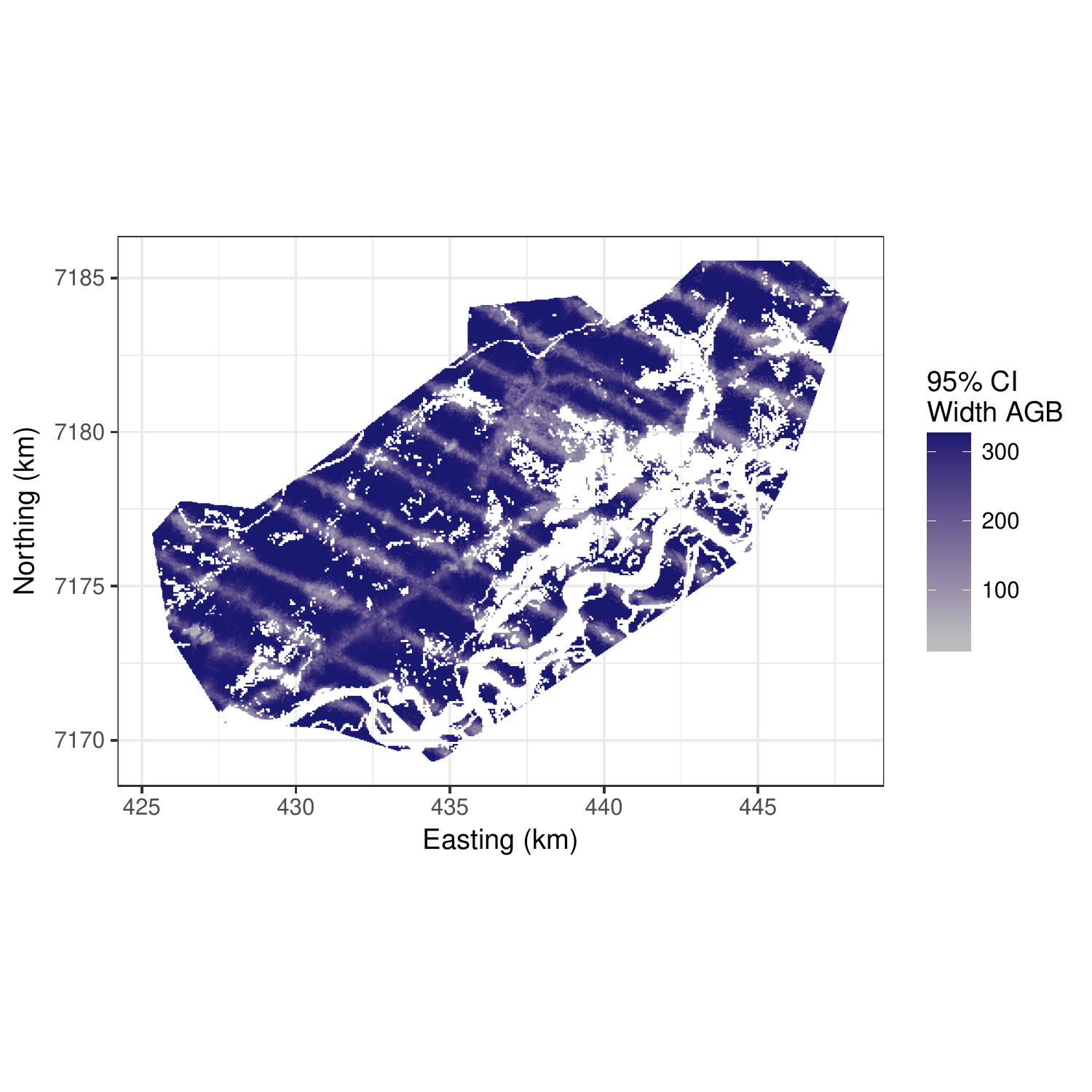}}\\
\subfigure[TD Median (1000 trees ha$^{-1}$)]{\includegraphics[scale=0.45,trim={0cm 3.0cm 0.3cm 3cm},clip]{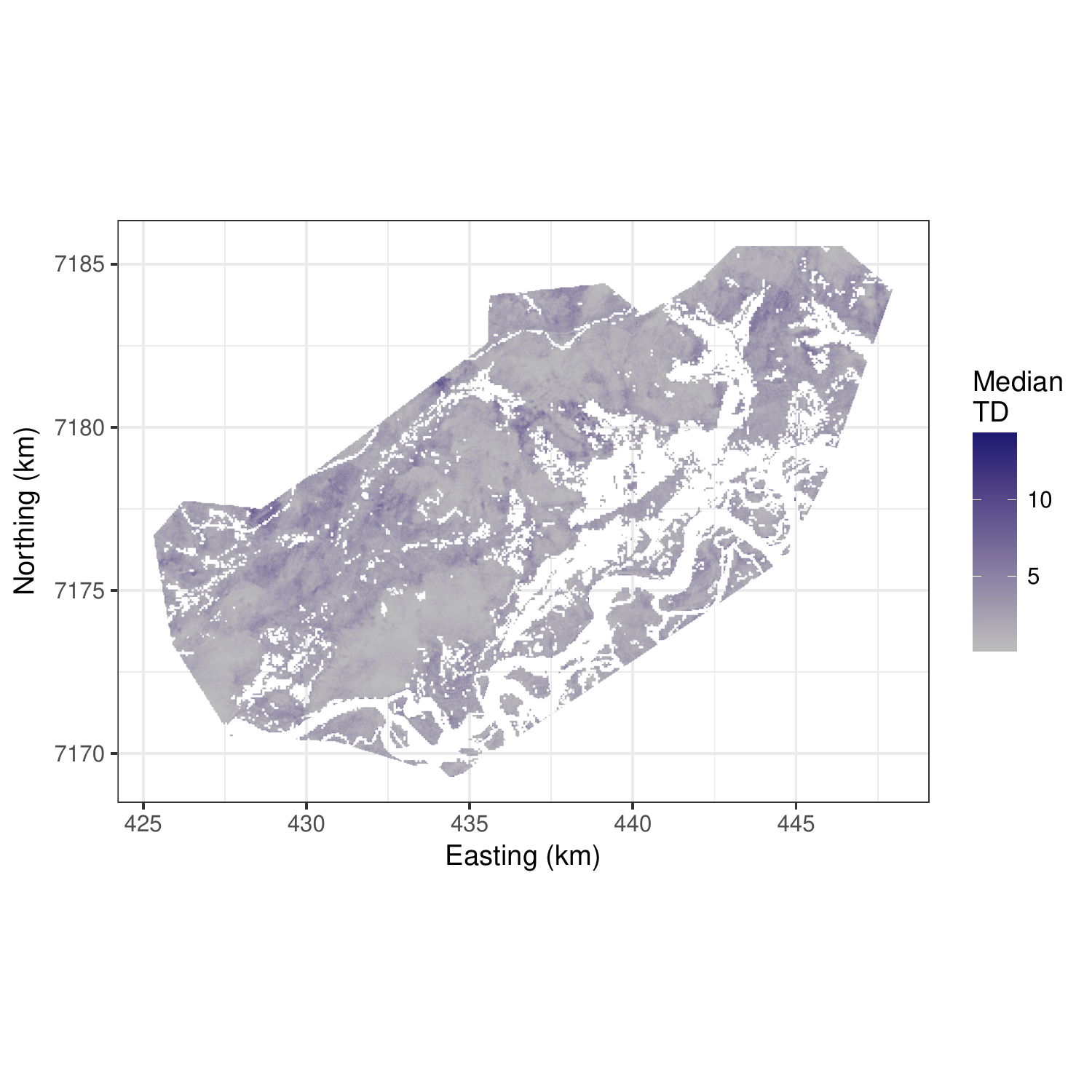}}
\subfigure[TD 95\% CI Width (1000 trees ha$^{-1}$)]{\includegraphics[scale=0.45,trim={0cm 3.0cm 0.3cm 3cm},clip]{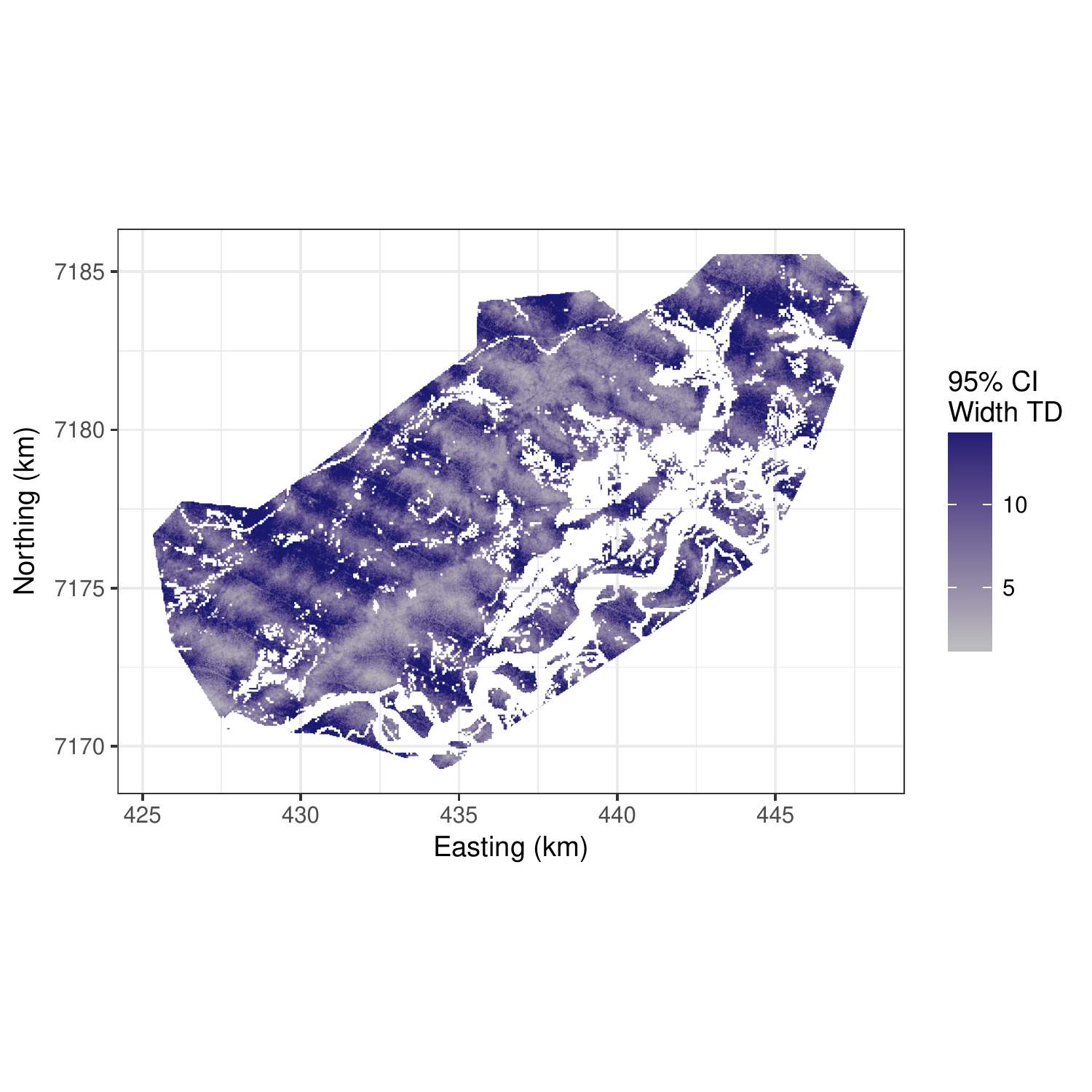}}\\
\subfigure[BA Median (m$^2$ ha$^{-1}$)]{\includegraphics[scale=0.45,trim={0cm 3.0cm 0.3cm 3cm},clip]{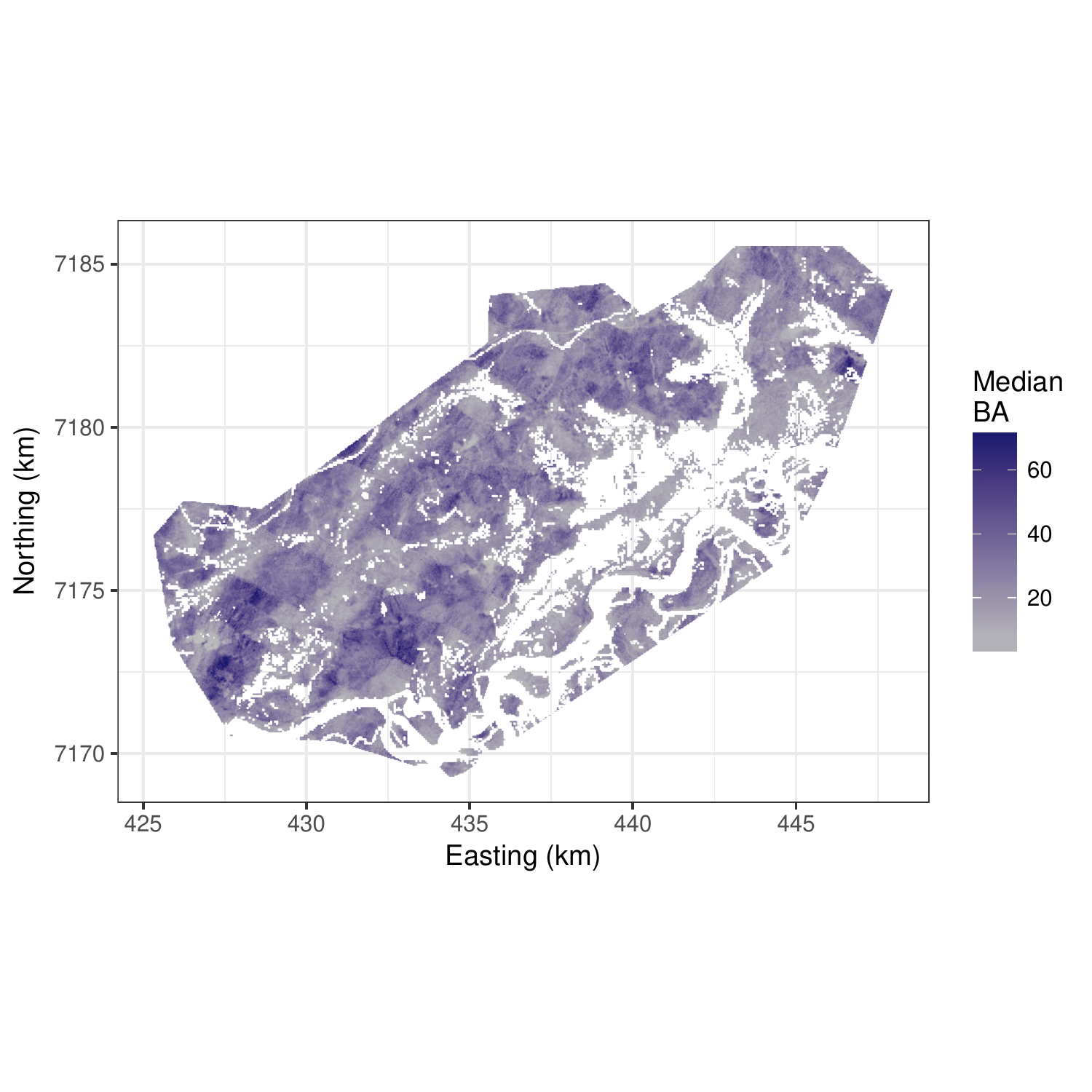}}
\subfigure[BA 95\% CI Width (m$^2$ ha$^{-1}$)]{\includegraphics[scale=0.45,trim={0cm 3.0cm 0.3cm 3cm},clip]{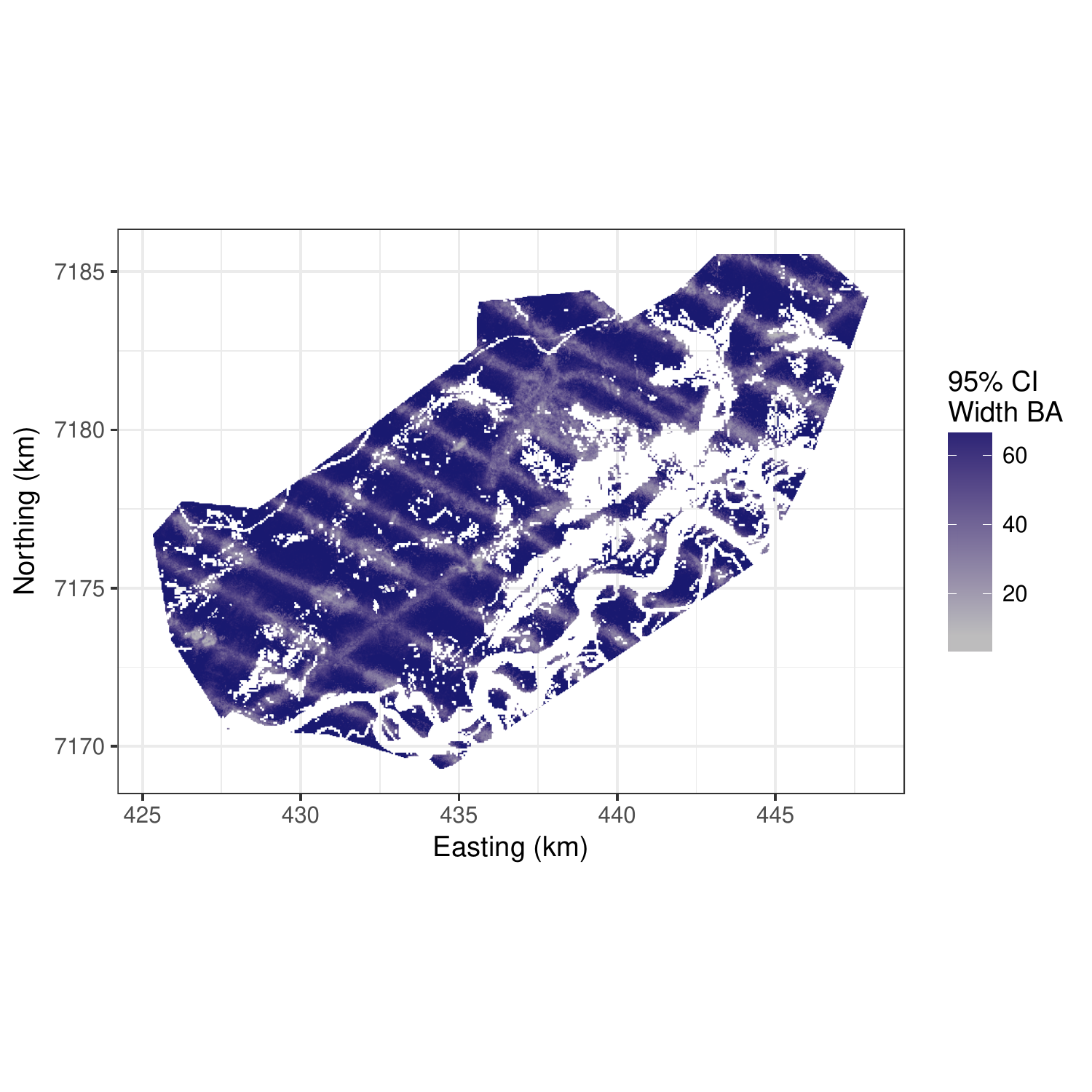}}
\caption{Model $q_w=5$ posterior predictive distribution median and 95\% CI width for AGB, TD, and BA forest variables over Bonanza Creek Experimental Forest.}\label{fig:pred-maps}
\end{figure}

As detailed in Section~\ref{sec:intro}, complete coverage maps of the forest outcomes with associated uncertainty estimates are important data products that can be delivered by the proposed two stage model. Following Section~\ref{ssec:postpred} and using the full data set depicted in Figure~\ref{fig:bc-map}, we predicted the forest outcomes on a 30$\times$30 m grid over the BCEF. Figure~\ref{fig:pred-maps} provides median and 95\% credible interval width maps for each outcome. Non-forested areas were omitted (white regions on the maps). Posterior predictive point estimates match well with the distribution of the forest outcomes across the BCEF and are clearly informed by the LiDAR factors which are capturing key forest structure characteristics. Most importantly, the prediction uncertainty maps, displayed in the right column of Figure~\ref{fig:pred-maps}, accurately reflect our lack of information for prediction units that are far from the flight lines were LiDAR data are available, i.e., we achieve more precise posterior predictive distributions along and adjacent to locations where LiDAR data are available. Far from the LiDAR flight lines prediction is only informed by the Landsat 8 tassel cap predictor variables, which in this study explained very little variability in the forest outcomes.

\section{Concluding Remarks}\label{sec:discussion}

We formulated an approach to model high-dimensional spatial data over a large set of locations, and developed an efficient implementation in \textsf{C}$++$. The SF-NNGP enables the analysis of multivariate spatially referenced datasets that, due to their magnitude, could not have been rigorously explored before. It does so by combining the ability of SFMs to compress the signal from high-dimensional structures into a few dimensions with the computational scalability of NNGPs.   

The algorithm was used to exploit the information from the high-dimensional LiDAR signals to jointly model and generate LiDAR based maps of multiple forest variables. Importantly, the proposed two stage model provides a viable approach to producing spatially continuous maps from sparsely sampled LiDAR and forest measurements, and delivers spatially explicit uncertainty quantification that captures the irregular distribution of information across the domain of interest. Such frameworks will become increasingly importantly as sampling LiDAR systems, such as GEDI, come on-line in the near future. These approaches can also be extended to help guide LiDAR and field data acquisition to minimize prediction uncertainty. 

Importantly, when fitting a spatial factor model one must choose the number of factors $q_w$ to be used in the model; there are different strategies to address this issue.  The approach we adopt here --looking at the out-of-sample evaluation metrics for different choices of $q_w$ and selecting the one where the curves flatten out-- is a pragmatic solution and is similar in spirit to cross-validation approaches commonly used to tune hyper-parameters in richly parametrized models. Like any other cross-validation approach, this leads to additional computation, but parallel computing opens the possibility of conducting simultaneous MCMC runs for different values of $q_w$.  As shown, both in the simulation experiment as well as in the BCEF data analysis, this heuristic provides sufficiently good results. 
Other automated rank selection schemes are available in the literature, such as those proposed in \citet{Lopes2004} and in \citet{Ren2013}; however, these drastically increase the computational burden of an already computationally costly problem.  

A research direction we are keen on exploring is an extension for spatio-temporal data. For this type of data it is necessary to posit a strategy to select the neighbors in the spatio-temporal domain, following the discussion presented in \citet{Datta2016c}. 

Although our method presents a substantial improvement in terms of scalability over existing approaches, further efforts are required to scale multivariate spatial methods to truly massive datasets. For instance, the ultimate goal for forest variable mapping assisted by sampled LiDAR in interior Alaska is a complete-coverage map of the entire domain (e.g., 46 million ha), which could easily require models capable of assimilating LiDAR signals in more than $10^{8}$ locations.

\vskip 14pt
\noindent {\large\bf Supplementary Materials}

The supplementary materials include (1) background information on NNGPs and spatial factor models, (2) the sampling algorithm for the SF-NNGP, and (3) additional simulation results.

\par

\vskip 14pt
\noindent {\large\bf Acknowledgments}

The research presented in this study was partially supported by NASA's Arctic-Boreal Vulnerability Experiment (ABoVE) and Carbon Monitoring System (CMS) programs. Additional support was provided by the United States Forest Service Pacific Northwest Research Station. Finley was supported by National Science Foundation (NSF) DMS-1513481, EF-1137309, EF-1241874, and Finley and Taylor-Rodriguez were supported on EF-1253225. Banerjee was supported by NSF DMS-1513654 and NSF IIS-1562303.

\par

\markboth{\hfill{\footnotesize\rm Daniel Taylor-Rodriguez AND Andrew Finley AND Abhirup Datta AND Chad Babcock AND Hans-Erik Andersen AND Bruce D. Cook AND Douglas C. Morton AND Sudipto Baneerjee} \hfill}
{\hfill {\footnotesize\rm Spatial Factor Models for Massive Datasets} \hfill}

\bibhang=1.7pc
\bibsep=2pt
\fontsize{9}{14pt plus.8pt minus .6pt}\selectfont
\renewcommand\bibname{\large \bf References}
\expandafter\ifx\csname
natexlab\endcsname\relax\def\natexlab#1{#1}\fi
\expandafter\ifx\csname url\endcsname\relax
  \def\url#1{\texttt{#1}}\fi
\expandafter\ifx\csname urlprefix\endcsname\relax\def\urlprefix{URL}\fi

\bibliographystyle{apalike}

\vskip .65cm
\noindent
Department of Mathematics \& Statistics, Portland State University, Portland, OR

\noindent
E-mail: dantayrod@pdx.edu
\vskip 2pt

\noindent
Department of Forestry, Michigan State University, East Lansing, MI

\noindent
E-mail: finleya@msu.edu
\vskip 2pt

\noindent
Department of Biostatistics, Johns Hopkins University, Baltimore, MA

\noindent
E-mail: abhidatta@jhu.edu
\vskip 2pt

\noindent
School of Environmental and Forest Sciences, University of Washington, Seattle, WA

\noindent
E-mail: babcoc76@uw.edu
\vskip 2pt

\noindent
USDA Forest Service Pacific Northwest Research Station, Seattle, WA

\noindent
E-mail: handersen@fs.us
\vskip 2pt

\noindent
Biospheric Sciences Laboratory, NASA Goddard Space Flight Center, Greenbelt, MD

\noindent
E-mail: bruce.cook@nasa.gov
\vskip 2pt

\noindent
Biospheric Sciences Laboratory, NASA Goddard Space Flight Center, Greenbelt, MD

\noindent
E-mail: douglas.morton@nasa.gov

\vskip 2pt
\noindent
Department of Biostatistics, University of California Los Angeles, Los Angeles, CA

\noindent
E-mail: sudipto@ucla.edu

\appendix
\pagebreak 

\section{Background Information}

\subsection{Dimension reduction with factor models} \label{subsec:sfm}

For a general $l$-variate problem, where the dependence is specified through a Gaussian Process $\bw^\star(\bs)\sim \text{GP}_l(\0,\mcal{C}(\cdot\given \btheta))$, the specification of $\mcal{C}(\cdot|\btheta)$ is complicated by the fact that it must be a nonnegative definite function, and it must meet the symmetry constraint $\mcal{C}(\bh\given\btheta)=\mcal{C}(-\bh\given\btheta)$ \citep[see][]{ver1998modeling,chiles2009geostatistics,Genton2015}.  In addition to the difficulty in specifying a suitable covariance function, the size of the LiDAR signals in our application makes modeling directly this joint high-dimensional spatial component a computationally daunting task.  As such, we take advantage of the gains achieved using the spatial factor model structure, which reduces the computational burden in two ways.  First, it dramatically reduces the dimensionality of the stochastic processes used, and second, it assumes that the multivariate stochastic processes considered are composed of independent univariate processes.  

Under the SFM structure, the spatial dependence is introduced by defining the spatial process as $\bw^\star(\bs)=\bLambda\bw(\bs)\sim\text{GP}(\0,\mcal{H}(\cdot \given \bphi))$, where  $\bLambda$ is a factor loadings matrix (commonly tall and skinny) and $\bw(\bs)$ is a small-dimensional vector of independent spatial GP's, providing the non-separable multivariate cross-covariance function given by 
\bea
\mcal{H}(\bh\given\bphi)&=&\text{cov}(\bLambda\,\bw(\bs), \bLambda\,\bw(\bs+\bh))\nonumber\\
&=& \sum_{k=1}^q \rho_k(\bh,\phi_k) \blambda_k \blambda_k^\prime \; = \;  \sum_{k=1}^q \mcal{C}_k(\bh,\phi_k) T_k, \label{covmatsup}
\eea
for locations $\bs,\bs+\bh\in \mcal{D}$.  Here, $\mcal{C}_k(\bh|\phi_k)$'s are univariate parametric correlation functions, and $\blambda_k$ is the $k$th column of $\bLambda$, which also corresponds to the eigenvector associated to the \emph{only} positive eigenvalue of the rank one matrix $T_k$. This cross-covariance matrix is induced by $q$-variate ($q\leq l$) spatial factors $\bw(\bs)$ with \emph{independent} components $w_k(\bs)\sim \text{GP}(0,\mcal{C}_k(\cdot\given \phi_k))$.  Hence, $\text{var}(w_k(\bs))=1$, $\text{cov}(w_k(\bs),w_{r}(\br))=0$ for $k\not=r$, and $\text{cov}(w_k(\bs),w_k(\bs+\bh))=\mcal{C}_k(\bh\given\phi_k)$. 


Additional constraints are required for factor models to be identifiable \citep{anderson2003}.  Nevertheless, conveniently with spatial factor models only two groups of orthogonal transformations lead to non-identifiability issues, as shown in \citet{Ren2013} .  The first of them is produced by an orthogonal matrix  $\bP_H$ resulting from the product of Householder reflectors, which is diagonal with $1$'s and $-1$'s.  The non-identifiability comes from the fact that $\bP_H^\prime \mcal{H}(\bh\given\bphi)\bP_H^\prime=\mcal{H}(\bh\given\bphi)$. The second type are permutation matrices $\bP_P$, given that $\bLambda \bP_P \bP_P^\prime \bw(\bs)\overset{d}{=} \bLambda \bw(\bs)$, where ``$\overset{d}{=}$'' represents equality in distribution. Both of these situations can be avoided, either through the conventional approach of making the upper triangle of the loadings matrix equal to 0 and its diagonal elements all equal to 1; or as in \citet{Ren2013}, by fixing the sign of one element in each column of $\bLambda$, while enforcing an ordering constraint on the univariate correlation functions. 

\subsection{Nearest Neighbor Gaussian Processes}\label{subsec:nngp}

In spite of the dimension reduction achieved with the factor model structure, given the formidable number of locations considered, even the factor model representation is prohibitive with dense Gaussian processes. Under a Bayesian approach, $q_w+q_v$ covariance matrices, each of dimension $n\times n$, have to be estimated and inverted at each iteration of the sampling algorithm. In view of this, we resort to the sparse approximation provided by the NNGP approach.

The Nearest Neighbor Gaussian Process approach belongs to the class of sparsity inducing methods that introduce zeros in the precision matrix to impose conditional independence, exploiting the graphical structure available for points distributed across space and/or time. The idea underlying this method is to derive a sparse approximation of a parent GP, which is a proper GP itself. The NNGP has been shown to provide an accurate and computationally efficient approximation to the dense parent GPs \citep[see for example][]{Datta2016,Datta2016a,Datta2016c,Finley2017}.

To elaborate, consider a univariate spatial Gaussian Process $w(\bs)\sim\text{GP}(0,\mcal{C}(\cdot \given\bphi))$ for $\bs\in\mcal{D} \subset\mathbb{R }^d$.  Recall that when observed at a finite collection of locations $\oset=\lrb{\bs_1,\ldots,\bs_n}$,  the process constrained to these locations is such that $\bw=(w(\bs_1),\ldots,w(\bs_n))'\sim \tN_n\lrp{\0,\bC}$, with $\bC=\Big(\Big(\mcal{C}(\|\bs_i-\bs_j\|;\bphi) \Big)\Big)$.  Alternatively, this joint density can be decomposed into the product of conditionals
\[p\lrp{w_1}p\lrp{w_2 | w_1}\ldots p\lrp{w_i | w_j: 1\leq j <i } \ldots p\lrp{w_n | w_j: 1\leq j <n },\]  
where $w_i=w(\bs_i)$. This representation and the multivariate normality of $\bw$ imply the linear model given by  $w_1=\eta_1$ and  $w_i= \sum_{j=1}^{i-1}b_{ij} w_{j} + \eta_i$,  with $\eta_i\sim N(0,\tau_{i})$, where $\tau_1=\text{var}(w_1)$ and $\tau_{i}=\text{var}(w_i | w_j: 1\leq j<i)$. 

If locations are suitably ordered, a good approximation can be obtained by replacing the conditioning set $\lrb{w_j: 1\leq j <i}$, for $i=2,\ldots,n$, by a subset $N(i)$ that contains a reduced number of nearest neighbors. When considering neighborhoods of sizes up to $m$, the sparse approximation to the dense linear model becomes  
\bea
w_i &=& \sum_{\bs_j\in N(i)}b_{ij} w_{j} + \xi_i 
\;=\; \bb_i' \bw_{N(i)} +  \xi_i, \label{eq:nngplm}
\eea
where $N(i)$ contains the $m_i=\text{min}\lrb{m,i-1}$ nearest neighbors within the conditioning set $\lrb{w_j: 1\leq j <i}$, and $\bw_{N(i)}=\lrb{w_j: \bs_j\in N(i)}$. Denote by $\bC_{U,V}$ represent the submatrix of $\bC$ indexed by the rows corresponding to locations in set $U$ and by columns indexed by locations in $V$, and $\bC_U$ be the square matrix with rows and columns indexed by locations indexed by $U$.  Using this notation,  we have that $\bb_i'=\bC_{i,N(i)}\bC_{N(i)}^{-1}$, i.e., the kriging weights conditioned on the neighbor set.  Additionally, the last term on the right hand side $\xi_i\sim N(0,F_{i})$, where $F_{i}=\text{var}(w_i | \bw_{N(i)})=\bC_{i}-\bC_{i,N(i)} \bC_{N(i)}^{-1}\bC_{N(i),i}$. Hence, both $\bb_i$ and $F_i$ are entirely characterized by the covariance function $\mcal{C}(\cdot,\cdot\given \bphi)$ from the parent process.  Note that the dense and the sparse process share the same equations for locations $i = 1,2,\ldots, m+1$.  This implies that the covariance among the first $m+1$ locations under the NNGP is the same as that of the parent process.

In vector form, the sparse model can be written as $\bw = \bB \bw + \mbs{\xi}$. Here,  $\mbs{\xi} \sim \tN_n(\0,\bF)$, where $\bF=\text{diag}\lrb{F_i:i=1,\ldots,n}$, and $\bB$ is the lower triangular matrix with zeros along the diagonal, and at most $m$ nonzero values in each row.  The nonzero values in the $i$th row of $\bB$ are located in columns $\lrb{j:\bs_j\in N(i)}$.  Hence, this representation implies that 
\be
\bw = (\tI-\bB)^{-1} \mbs{\xi} \sim \tN_n(\0,\tbC),
\ee
where $\tbC=(\tI - \bB)^{-1}\bF (\tI - \bB)^{-T}$ (with $\tbC^{-1}$ sparse), which provides a good approximation to the original covariance matrix $\bC$.  For more details on the construction and appealing features of the NNGP methodology, we direct the reader to the meticulous construction presented in \citet{Datta2016c}.

\section{Sampling algorithm}\label{app:sampling}

To begin with, given that we have a two-step process, where $\bw(\bs)$ is assumed to exclusively capture the spatial patterns in the $\bz(\bs)$, the full conditional for $\bw(\bs_i)$ is proportional to
\bea
\tN_{q_w}\lrp{\bw(\bs_i) | \bB_i^{(w)} \bw_{N(i)},\bF_i^{(w)} } \prod_{\bs_j\in\parent(i)} \tN_{q_w}\lrp{\bw(\bs_j) \given \bB_j^{(w)} \bw_{N(j)},\bF_j^{(w)} } \times{}\nonumber\\
\tN_{l}\lrp{\bz(\bs_i) \given \bX_z(\bs_i)^\prime\bbeta_z + \bLambda_z \bw(\bs_i), \bPsi_z},\label{eq:postWun}
\eea
where $\parent(i)=\lrb{\bs_j\in \oset: \bs_i \in N(j)}$, is the set of locations to which $\bs_i$ is a neighbor. To simplify \eqref{eq:postWun}, let $\bs_{j_d}$ be the $d$th neighbor of $\bs_j\in\mcal{D}$ (for $1\leq d\leq m$). The columns in $\bB_j^{(w)}$ indexed by the set \[\mcal{I}_{N(j),\bs_{j_d}}=\lrb{(d-1)q+1,\ldots,(d-1)q+q}\] relate $\bs_j$ and $\bs_{j_d}$. Denote by $B_{j,j_d}^{(w)}$ the $q\times q$ matrix containing the columns indexed by $\mcal{I}_{N(j),\bs_{j_d}}$ in $\bB_j^{(w)}$. From the component of the expression above corresponding to locations $\bs_j\in\parent(i)$, we may rewrite the quadratic form within the exponential function in the normal density, as
\bean
&&\Big(\bw(\bs_j)- \sum_{j_d\in N(j)} B_{j,j_d}^{(w)} \bw(\bs_{j_d})\Big)^\prime \bF_j^{-1} \Big(\bw(\bs_{j})- \sum_{j_d\in N(j)} B_{j,j_d}^{(w)}\bw(\bs_{j_d})\Big)\;=\\
&&\qquad \qquad \qquad\qquad \qquad \qquad\qquad \qquad \qquad \lrp{B_{j,i}^{(w)}\bw(\bs_i)- \bchi_{j,i}^{(w)}}^\prime (\bF_j^{(w)})^{-1} \lrp{B_{j,i}^{(w)}\bw(\bs_i)- \bchi_{j,i}^{(w)}},
\eean
where $\bchi_{j,i}^{(w)}=\bw(\bs_j) - \sum_{j_d \not=i} B_{j,j_d}^{(w)} \bw(\bs_{j_d})$.  

Making use of this notation, the full conditional for $\bw(\bs_i)$ for $\bs_i\in\refset$, is $\tN_{q_w}(\bw(\bs_i)| \bSigma_i^{(w)}\bmu_i^{(w)}, \bSigma_i^{(w)})$, with
{\footnotesize
\bean
\bmu_i^{(w)}&=&\lrp{(\bF_i^{(w)})^{-1}\bB_i^{(w)} \bw_{N(i)} + \sum_{\bs_j\in\parent(i)}(B_{j,i}^{(w)})^\prime (\bF_j^{(w)})^{-1}\bchi_{j,i}^{(w)}+\bLambda_z^\prime \bPsi_z^{-1}(\bz(\bs_i)-\bX_z(\bs_i)^\prime \bbeta_z) }\text{, and}\\
\bSigma_i^{(w)} &=& \lrp{(\bF_{i}^{(w)})^{-1} + \sum_{\bs_j\in\parent(i)}(B_{j,i}^{(w)})^\prime (\bF_j^{w)})^{-1}B_{j,i}^{(w)} +\bLambda_z^\prime \bPsi_z^{-1}\bLambda_z }^{-1}.
\eean}
Similarly, the full conditional for $\bv(\bs_i)$ is  $\tN_{q_v}(\bv(\bs_i)| \bSigma_i^{(v)}\bmu_i^{(v)}, \bSigma_i^{(v)})$, where
{\footnotesize
\bean
\bmu_i^{(v)}&=&\lrp{(\bF_i^{(v)})^{-1}\bB_i^{(v)} \bv_{N(i)} + \sum_{\bs_j\in\parent(i)}(B_{j,i}^{(v)})^\prime (\bF_j^{(v)})^{-1}\bchi_{j,i}^{(v)}+\bGamma^\prime \bPsi_y^{-1}(\by(\bs_i)-\bX_y(\bs_i)^\prime \bbeta_y - \bLambda_y \bw(\bs_i)) }\text{, and}\\
\bSigma_i^{(v)} &=& \lrp{(\bF_{i}^{(v)})^{-1} + \sum_{\bs_j\in\parent(i)}(B_{j,i}^{(v)})^\prime (\bF_j^{(v)})^{-1}B_{j,i}^{(v)} +\bGamma^\prime \bPsi_y^{-1}\bGamma  }^{-1}.
\eean}


To obtain the full conditional density for $\bbeta_z$ and $\bbeta_y$, let $\bW_{\oset_z}$ be the $n_z\times q_w$ matrix with rows given by $\bw(\bs)$ for $\bs\in\oset_z$. Define analogously the $n_y\times q_w$ and $n_y\times q_v$ matrices $\bW_{\oset_y}$  and $\bV_{\oset_y}$, respectively. Represent $\bLambda_z=(\blambda_1^z: \cdots:\blambda_{h_z}^z)^\prime$, $\bLambda_y=(\blambda_1^y: \cdots:\blambda_{h_y}^y)^\prime$, and $\bGamma=(\bgamma_1: \cdots:\bgamma_{h_y})^\prime$. Additionally, for $j=1,\ldots, h_z$ and $k=1,\ldots,h_y$, define
$\bz_{j}= (z_{j}(\bs):\bs\in \oset_z)'$,  and $\by_{k}= (y_{k}(\br):\br\in \oset_y)'$. Also denote the $n_z\times p_{z,j}$ matrix of predictors for the $j$th outcome in $\bz(\cdot)$ by $\bX_{j}^z=(\bx_{j}^z(\bs): \bs\in\oset_z)'$ for $j=1,\ldots,h_z$. Similarly, let $\bX_{k}^y=(\bx_{k}^y(\br_1):\ldots:\bx_{k}^y(\br_{n_y}))'$ for $k=1,\ldots,h_y$. Thus, assuming flat priors for $\bbeta_z$ and $\bbeta_y$ the full conditionals are
\bean
\bbeta_z\given \cdot &\sim& \prod_{j=1}^{h_z} \tN_{p_{z,j}}(\bbeta_{j}^z \given \Omega_j^z\bmu_j^z, \Omega_j^z),
\eean
where $\bmu_j^z = \lrp{\frac{1}{\psi_j^z}(\bX_j^z)^\prime (\bz_j-\bW_{\oset_z}\blambda_j^z) }$ and $\Omega_{j}^z=\psi_j^z\lrp{(\bX_j^z)^\prime \bX_j^z}^{-1}$, and 
\bean
\bbeta_y\given \cdot &\sim& \prod_{k=1}^{h_y} \tN_{p_{y,k}}(\bbeta_{k}^y \given \Omega_{k}\bmu_{k}, \Omega_{k}),
\eean
where $\bmu_k^y = \lrp{\frac{1}{\psi_{k}^y}(\bX_k^y)^\prime (\by_k-\bW_{\oset_y}\blambda_k^y -\bV_{\oset_y}\bgamma_k) }$ and $\Omega_{k}^y=\psi_k^y\lrp{(\bX_k^y)^\prime \bX_h}^{-1}$.

Given the identifiability restrictions imposed on $\bLambda_z$, let $q_j= \text{min}\lrb{j-1,q_w}$ for $2\leq j\leq h_z$, and denote by $\tilde{\blambda}_j^z=(\lambda_{j1}^z,\ldots, \lambda_{j q_j}^z)'$ the vector of unrestricted elements in the $j$th row of $\bLambda_z$.  Define $\bW_{1:j}=\lrp{\dot{\bw}_1 \cdots \dot{\bw}_j}$ (i.e., the matrix with the first $j$ columns of $\bW_{\oset_z}$).  Using the definitions above, the full conditional density for $\bLambda_z$ can be represented as $\prod_{j=2}^{h_z} \tN_{q_j}\lrp{\tilde{\blambda}_{j} \given \Omega_{\lambda_j^z}\bmu_{\blambda_j^z}, \Omega_{\lambda_j^z}}$, where 
\bean
\bmu_{\tilde{\blambda}_j^z} &=& \left\{ \bmat \frac{1}{\psi_j^z} \bW_{1:(j-1)}'(\bz_j-\bX_j^z\bbeta_j^z-\dot{\bw}_j)\hfill &\text{ if }&2\leq j\leq q_w  \\ 
 \frac{1}{\psi_j^z} \bW_{\oset_z}'(\bz_j-\bX_j^z\bbeta_j^z) \hfill &\text{ if }&j> q_w\emat\right.,\text{ and}\\
 &&\\
\Omega_{\tilde{\blambda}_j^z}&=&\left\{ \bmat \lrp{\frac{1}{\psi_j^z}\bW_{1:(j-1)}'\bW_{1:(j-1)}+\tI_{j-1}}^{-1} \hfill &\text{ if }&2\leq j\leq q_w  \\ 
  \lrp{\frac{1}{\psi_j^z}\bW_{\oset_z}'\bW_{\oset_z}+\tI_{q_w}}^{-1} \hfill &\text{ if }&j> q_w \emat\right.
\eean

The elements in $\bLambda_y$ are all unrestricted; hence, the full conditional posterior for $\bLambda_y$ corresponds to $\prod_{k=1}^{h_y} \tN_{q_w}\lrp{\blambda_{k}^y \given \Omega_{\lambda_{k}^y}\bmu_{\blambda_{k}^y}, \Omega_{\lambda_{k}^y}}$, with
\bean
\bmu_{\blambda_{k}^y}\;=\;\frac{1}{\psi_k^y} \bW_{\oset_y}'(\by_k-\bX_k^y\bbeta_k^y-\bV_{\oset_y}\bgamma_k),&\text{ and }&
\Omega_{\blambda_{k}^y}\;=\;\psi_k^y\lrp{\bW_{\oset_y}'\bW_{\oset_y}+\tI_{q_w}}^{-1}
\eean

The sampler for $\bGamma$ has a similar form to that of $\bLambda_z$, with the upper triangular elements equal to zero; however, given that we make no dimension reduction for the forest outcomes, the diagonal elements are only constrained to be positive (instead of setting them to one). First, for  $2\leq k\leq h_y$, let $q_k= \text{min}\lrb{k-1,q_v}$, denote $\tilde{\bgamma}_k=(\gamma_{k1},\ldots, \gamma_{k q_k})'$,  and let $\bV_{1:k}=\lrp{\dot{\bv}_1 \cdots \dot{\bv}_k}$ denote the matrix with the first $k$ columns of $\bV_{\oset_y}$.  Then, the full conditional posterior density for $\bGamma$ is $\prod_{k=2}^{h_y} \tN_{q_k}\lrp{\tilde{\bgamma}_{k} \given \Omega_{\gamma_k}\bmu_{\gamma_k}, \Omega_{\gamma_k}}$, with 
\bean
\bmu_{\tilde{\bgamma}_k} &=& \left\{ \bmat \frac{1}{\psi_k^y} \bV_{1:(k-1)}'(\by_k-\bX_k^y\bbeta_k^y-\bW_{\oset_y}\blambda_k^y-\gamma_{kk}\dot{\bv}_k)\hfill &\text{ if }&2\leq k\leq q_v  \\ 
 \frac{1}{\psi_k^y} \bV_{\oset_y}'(\by_k-\bX_k^y\bbeta_k^y-\bW_{\oset_y}\blambda_k^y) \hfill &\text{ if }& k> q_v\emat\right.,\text{ and}\\
 &&\\
\Omega_{\tilde{\bgamma}_k}&=&\left\{ \bmat \lrp{\frac{1}{\psi_k^y}\bV_{1:(k-1)}'\bV_{1:(k-1)}+\tI_{k-1}}^{-1} \hfill &\text{ if }&2 \leq k\leq q_v  \\ 
  \lrp{\frac{1}{\psi_k^y}\bV_{\oset_y}'\bV_{\oset_y}+\tI_{q_v}}^{-1} \hfill &\text{ if }& k > q_v \emat\right.
\eean

As mentioned before, the diagonal elements of $\bGamma$ are assumed to be positive.  Hence to sample the $k$th diagonal element in $\bGamma$ we assume a prior $\propto \tI_{\lrb{\gamma_{kk}>0}}$.  This yields a truncated normal distribution, obtained from a normal with mean $\mu_{\gamma_{kk}}$ and variance $\xi_{\gamma_{kk}}$ and truncated to be greater than zero, where 
\bean
\mu_{\gamma_{kk}} &=& (\dot{\bv}_r \dot{\bv}_k)^{-1} \dot{\bv}_r(\dot{\by}_k - \bX_k^y\bbeta_k^y -\bW_{\oset_y}\blambda_k^y -\bV_{1:(k-1)}\tilde{\bgamma}_k),  \\
\xi_{\gamma_{kk}}&=& \frac{\psi_k^y}{\dot{\bv}_r \dot{\bv}_k}.
\eean

Given that the half-$t$ distribution is a mixture of two Inverse-Gamma distributions, the full conditionals for $\bpsi_z$ and $\bpsi_y$ (the vectors of diagonal elements of $\bPsi_z$ and $\bPsi_y$, respectively) are conjugate with their corresponding likelihoods. Sampling them amounts to drawing from 
\bean
\bpsi_z|\cdot&\sim&\prod_{j=1}^{h_z} \mcal{IG}\lrp{\psi_j^z \,\Big|\,  \frac{\nu+n_z }{2},\frac{\nu}{a_j^z}+\frac{1}{2}\sum_{\bs_i\in \oset_z}(z_{j}(\bs_i)-\bx_{j}^z(\bs_i)'\bbeta_j^z -(\blambda_j^z)' \bw(\bs_i))^2},
\eean
with hyperparameters $\ba_{z}=(a_1^z,\ldots,a_{h_z}^z)$ sampled from 
\bean
\ba_{z}  | \bpsi_z &\sim & \prod_{j=1}^{h_z} \mcal{IG}\lrp{a_{j}^z \,\Big|\, \frac{1}{2},\frac{\nu}{\psi_j^z}+\frac{1}{A^2}}
\eean

\bean
\bpsi_y|\cdot&\sim&\prod_{k=1}^{h_y} \mcal{IG}\lrp{\psi_k^y \,\Big|\,  \frac{\nu+n_y }{2},\frac{\nu}{a_k^y}+\frac{1}{2}\sum_{\bs_i\in \oset_y}(y_k(\bs_i)-\bx_k^y(\bs_i)'\bbeta_k^y -(\blambda_k^y)' \bw(\bs_i) -(\bgamma_k)' \bv(\bs_i))^2},
\eean
with hyperparameters $\ba_{y}=(a_1^y,\ldots,a_{h_y}^y)$ sampled from 
\bean
\ba_{y}  | \bpsi_y &\sim & \prod_{k=1}^{h_y} \mcal{IG}\lrp{a_k^y \,\Big|\, \frac{1}{2},\frac{\nu}{\psi_k^y}+\frac{1}{A^2}}
\eean

Lastly, we may use a Metropolis within Gibbs steps to sample $\bphi_w$ and $\bphi_v$, with target densities proportional to
\bean
\pi(\bphi_w) \tN_{nq_w}(\bw_{\oset_z} | \0, \tbC^{(w)})  &=& \prod_{k=1}^{q_w} \pi(\phi_{w,k}) \tN_n\lrp{\dot{\bw}_k | \0 , \tbC_{k}^{(w)}(\phi_{w,k}) },\text{ and}\\
\pi(\bphi_v) \tN_{nq_v}(\bv_{\oset_y} | \0, \tbC^{(v)})  &=& \prod_{r=1}^{q_v} \pi(\phi_{v,r}) \tN_n\lrp{\dot{\bv}_{r} | \0 , \tbC_{r}^{(v)}(\phi_{v,r}) },\label{eq:postphi} 
\eean
which can both be sampled using a random walk Metropolis.  

\newpage
\section{Additional results from the simulation exercise}

\begin{figure}[H]
\centering
\subfigure[$q_w=3$]{\includegraphics[scale=0.5,trim={0 0.8cm 0 0.8cm},clip]{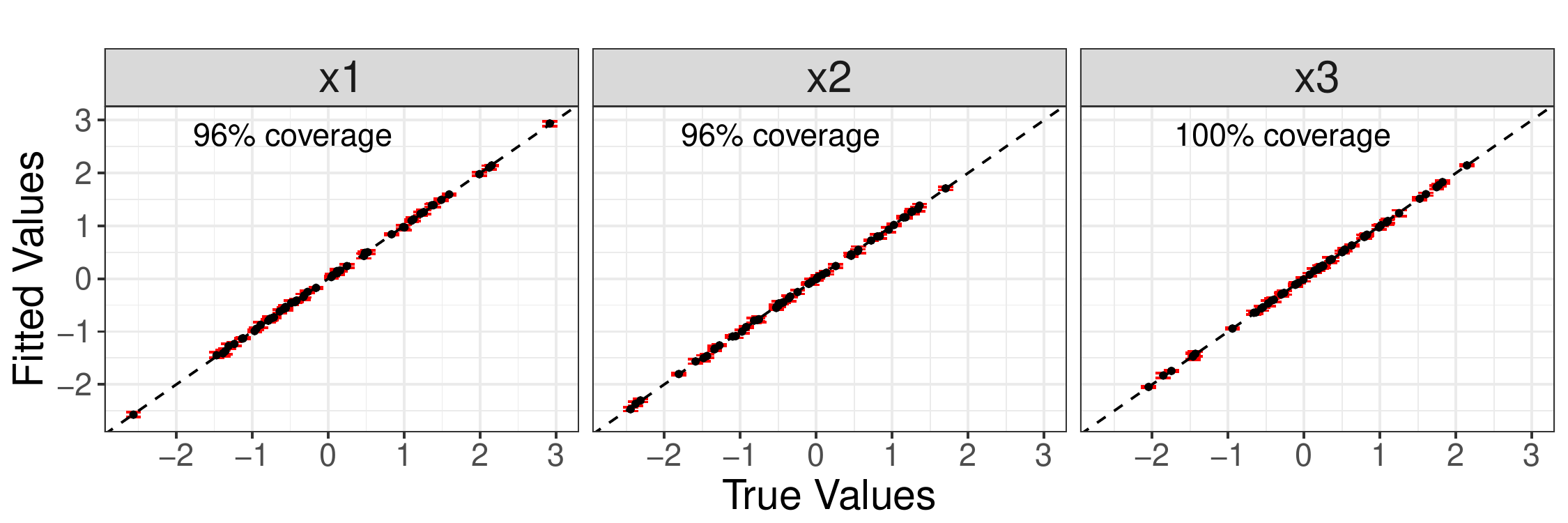}}
\subfigure[$q_w=5$]{\includegraphics[scale=0.5,trim={0 0.8cm 0 1.5cm},clip]{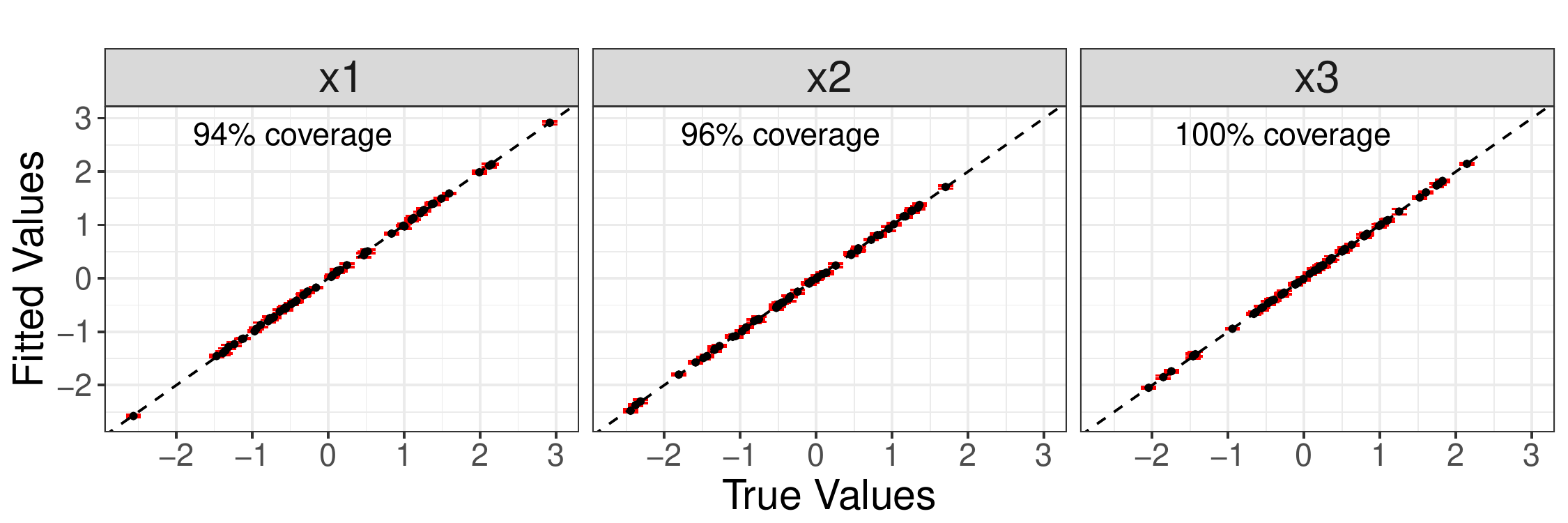}}
\subfigure[$q_w=8$]{\includegraphics[scale=0.5,trim={0 0.8cm 0 1.5cm},clip]{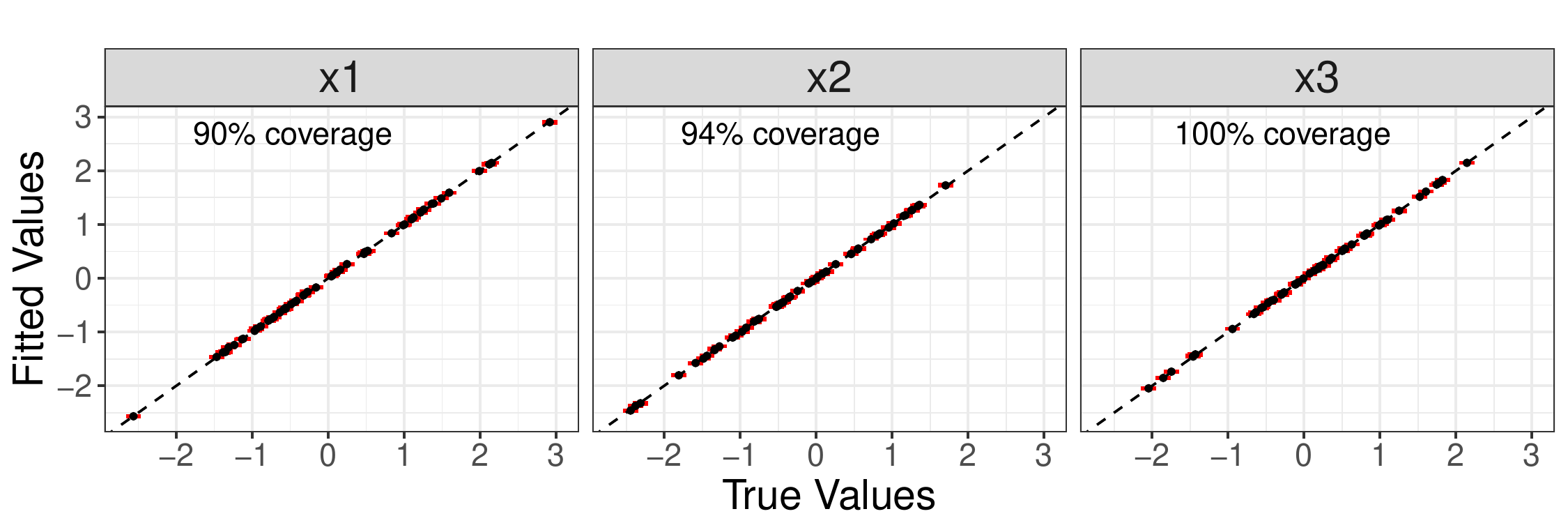}}
\subfigure[$q_w=10$]{\includegraphics[scale=0.5,trim={0 0.1cm 0 1.5cm},clip]{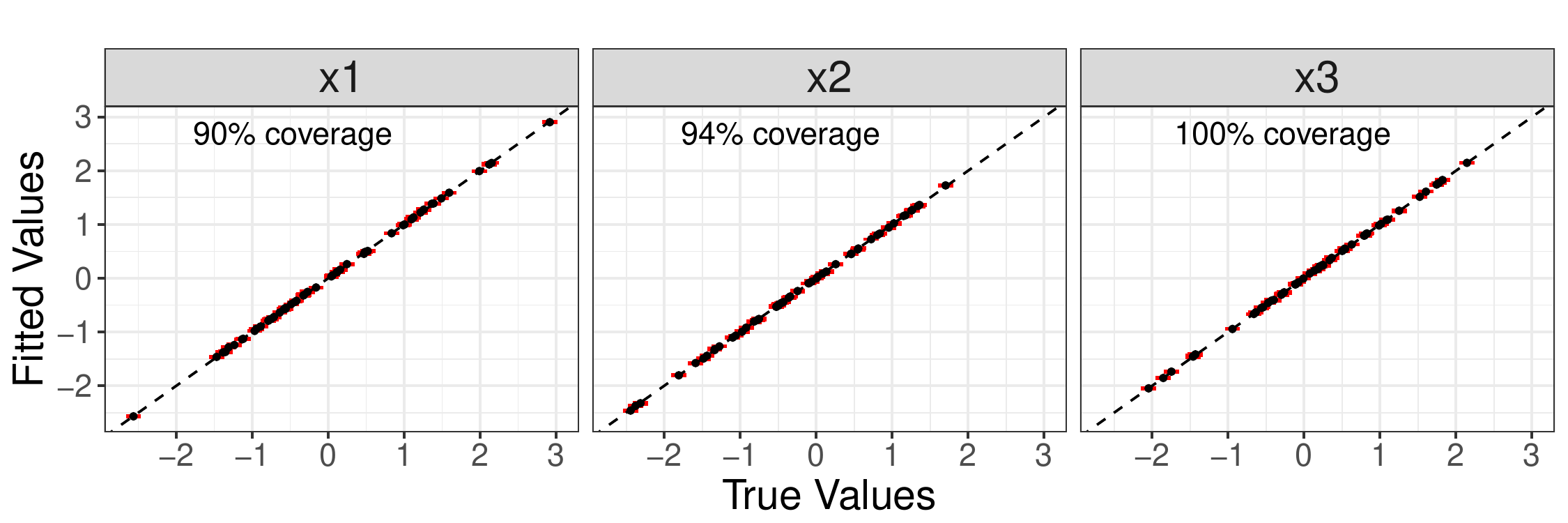}}
\caption{Posterior median and 95\% credible set vs true values of $\bbeta_z$ for $q_w\in\lrb{3,5,8,10}$. The rows vary by the number of spatial factors used in model fitting, the columns vary by predictor.}\label{fig:sim1betas1}
\end{figure}

\begin{figure}[H]
\centering
\subfigure[$q_w=3$]{\includegraphics[scale=0.28,trim={0 0 0 0.6cm},clip]{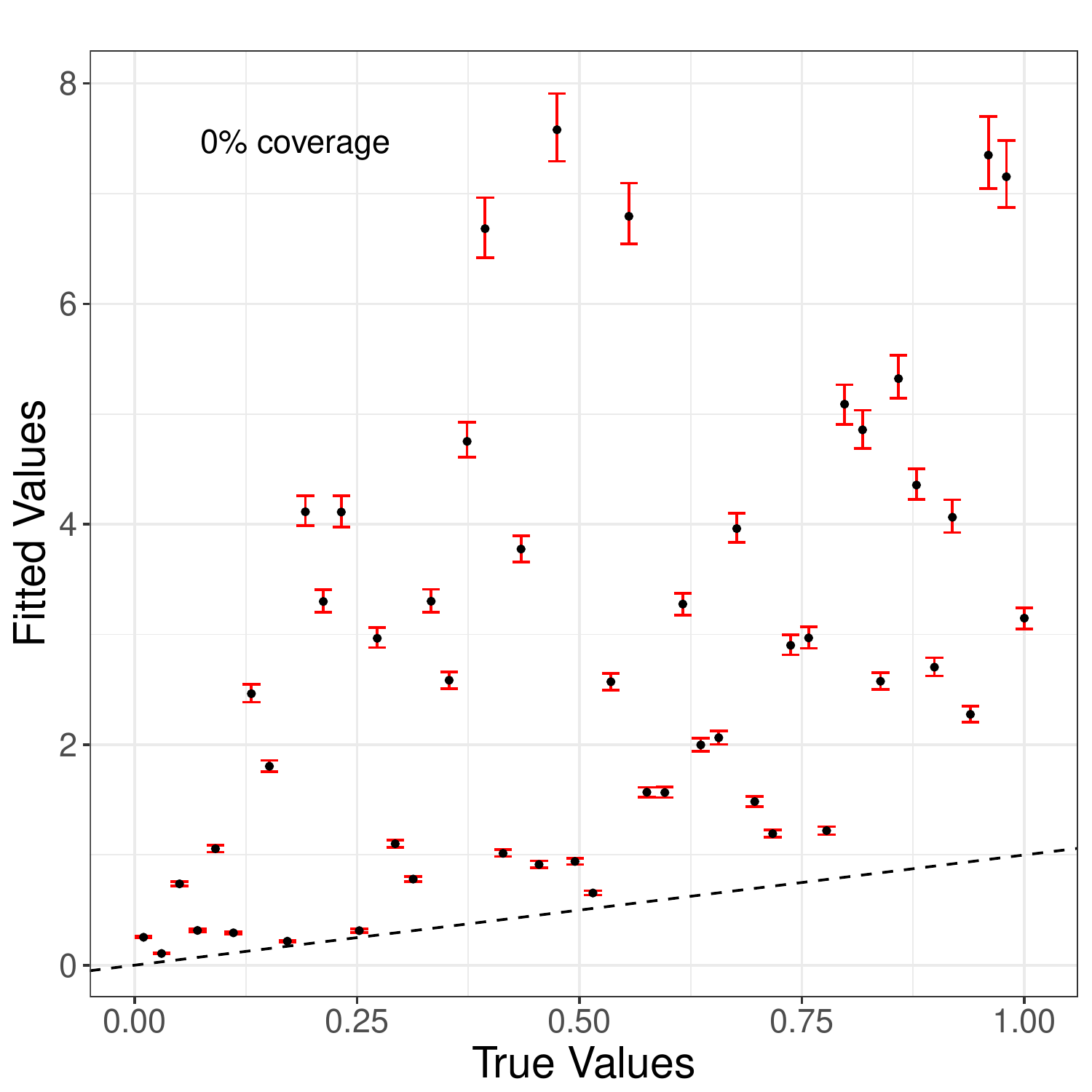}}
\subfigure[$q_w=5$]{\includegraphics[scale=0.28,trim={0.8cm 0 0 0.6cm},clip]{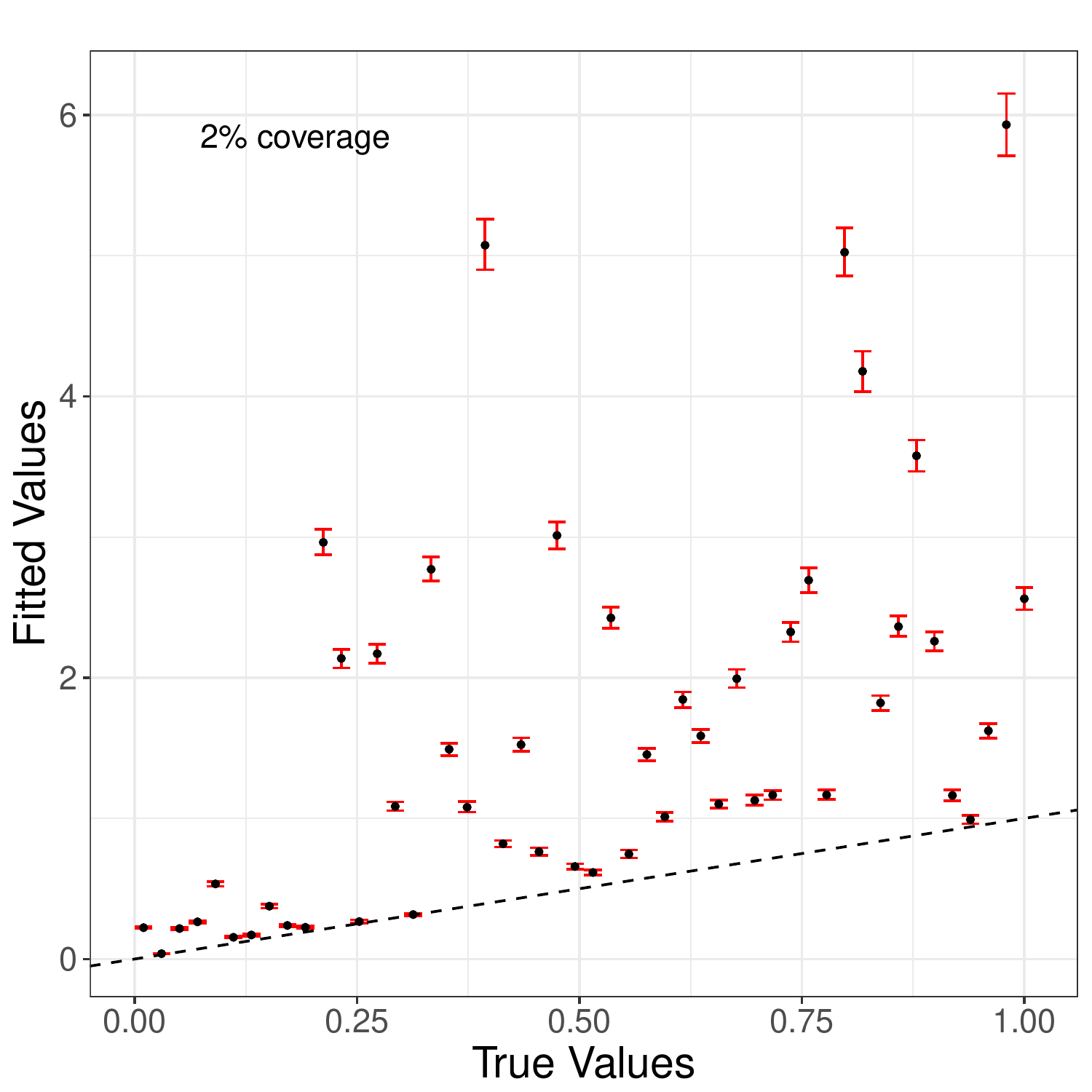}}\\
\subfigure[$q_w=8$]{\includegraphics[scale=0.28,trim={0 0 0 0.6cm},clip]{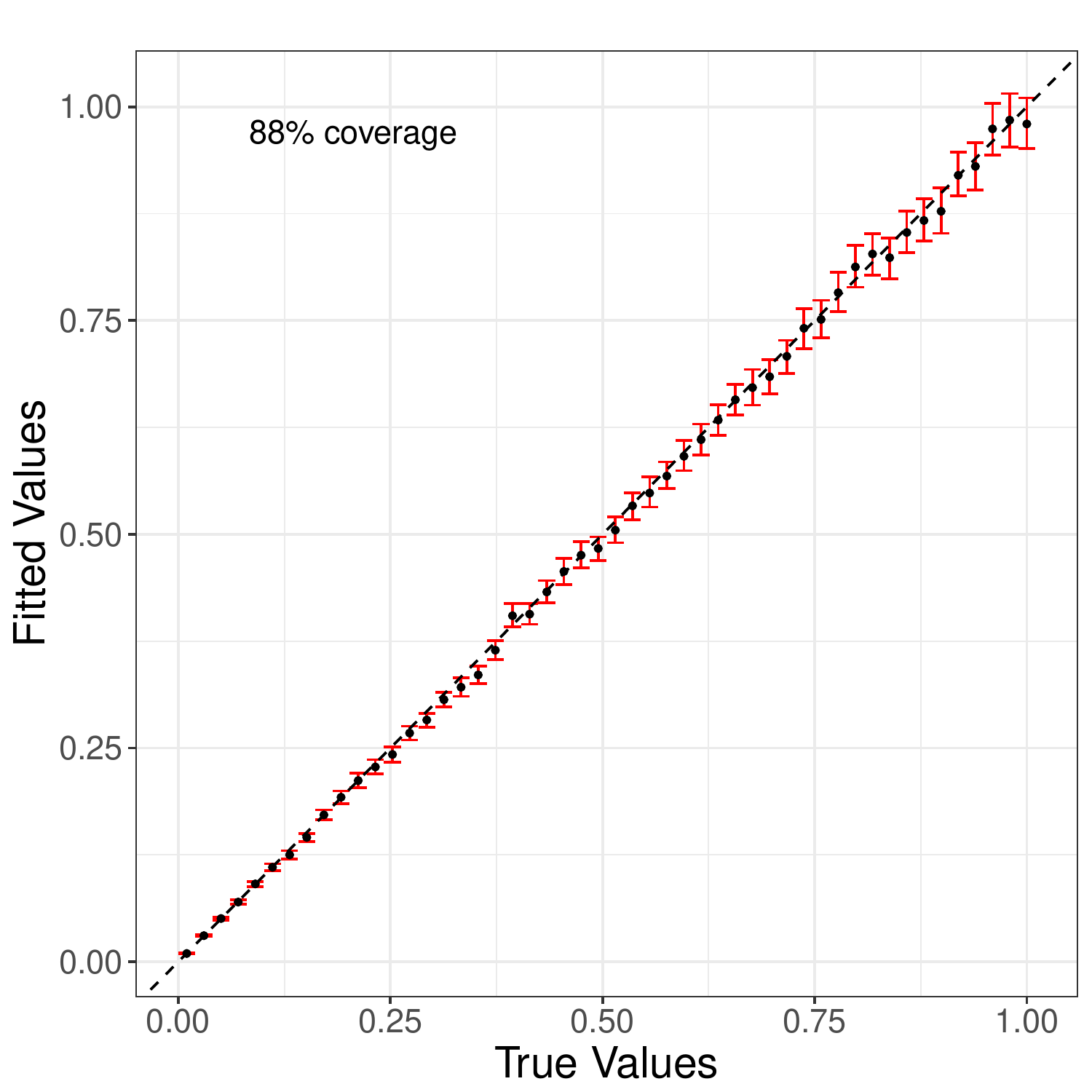}}
\subfigure[$q_w=10$]{\includegraphics[scale=0.28,trim={0.6cm 0 0 0.6cm},clip]{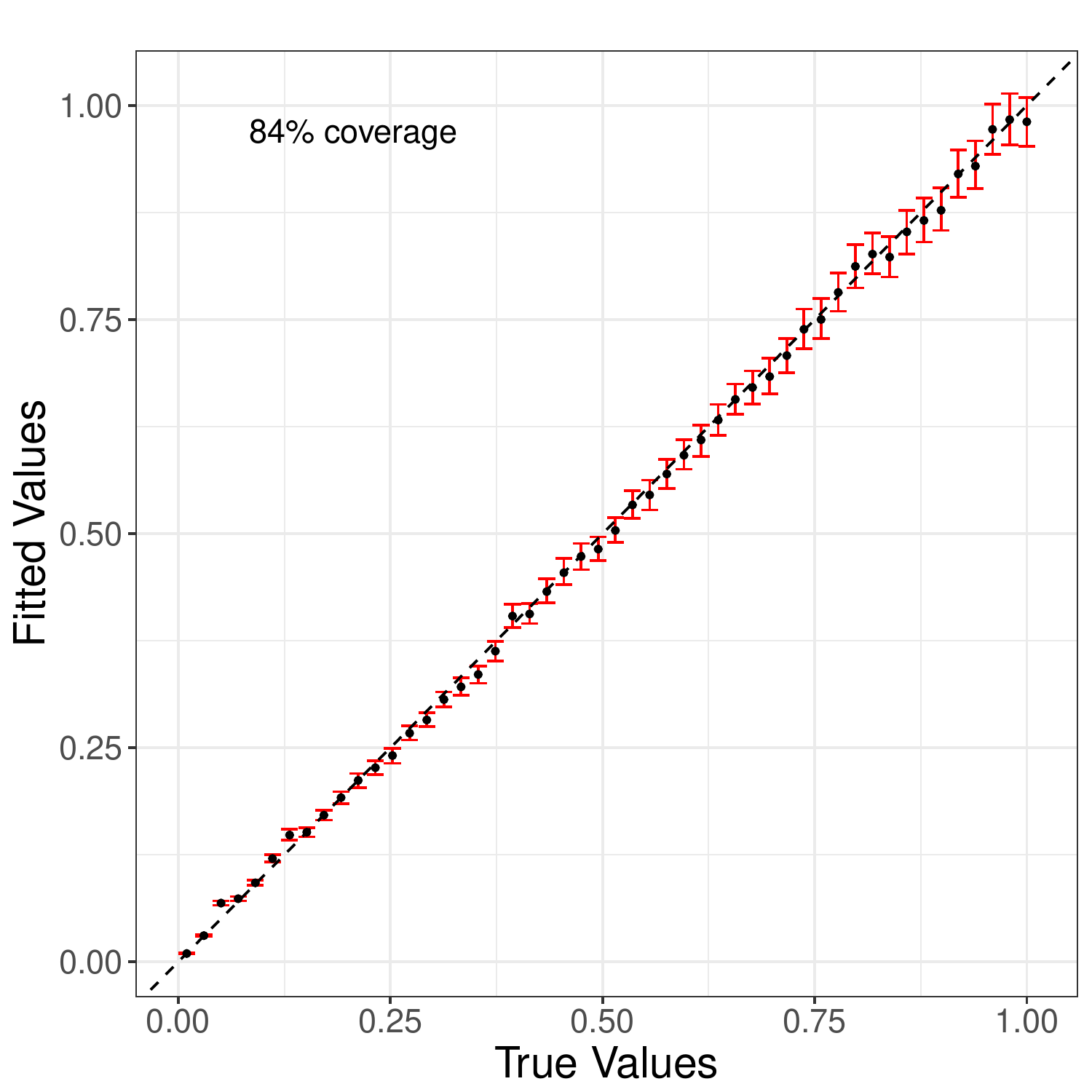}}
\caption{Posterior median and 95\% credible set vs true values of $\bpsi_z$.}\label{fig:sim1taus}
\end{figure}

\begin{sidewaysfigure}
\centering
\includegraphics[scale=0.5,trim={0 0 0 0.7cm},clip]{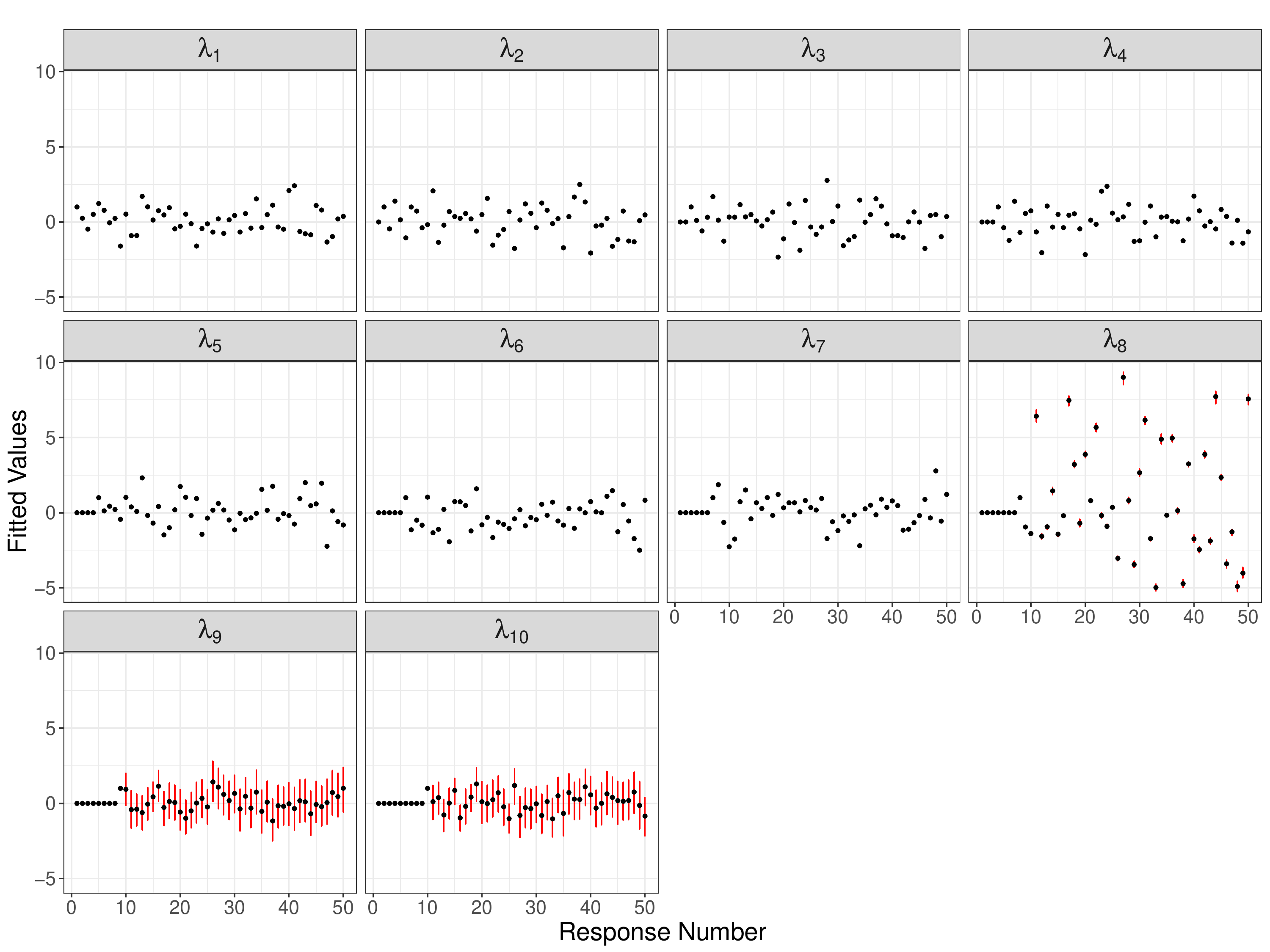}
\caption{Posterior median and 95\% credible set for the factor loadings matrix with $q_w=10$. Each panel displays a column of the estimated $\bLambda_z$ matrix.}\label{fig:sim1loadq10}
\end{sidewaysfigure}

\begin{sidewaysfigure}
\centering
\includegraphics[scale=0.5,trim={0 0 0 0.7cm},clip]{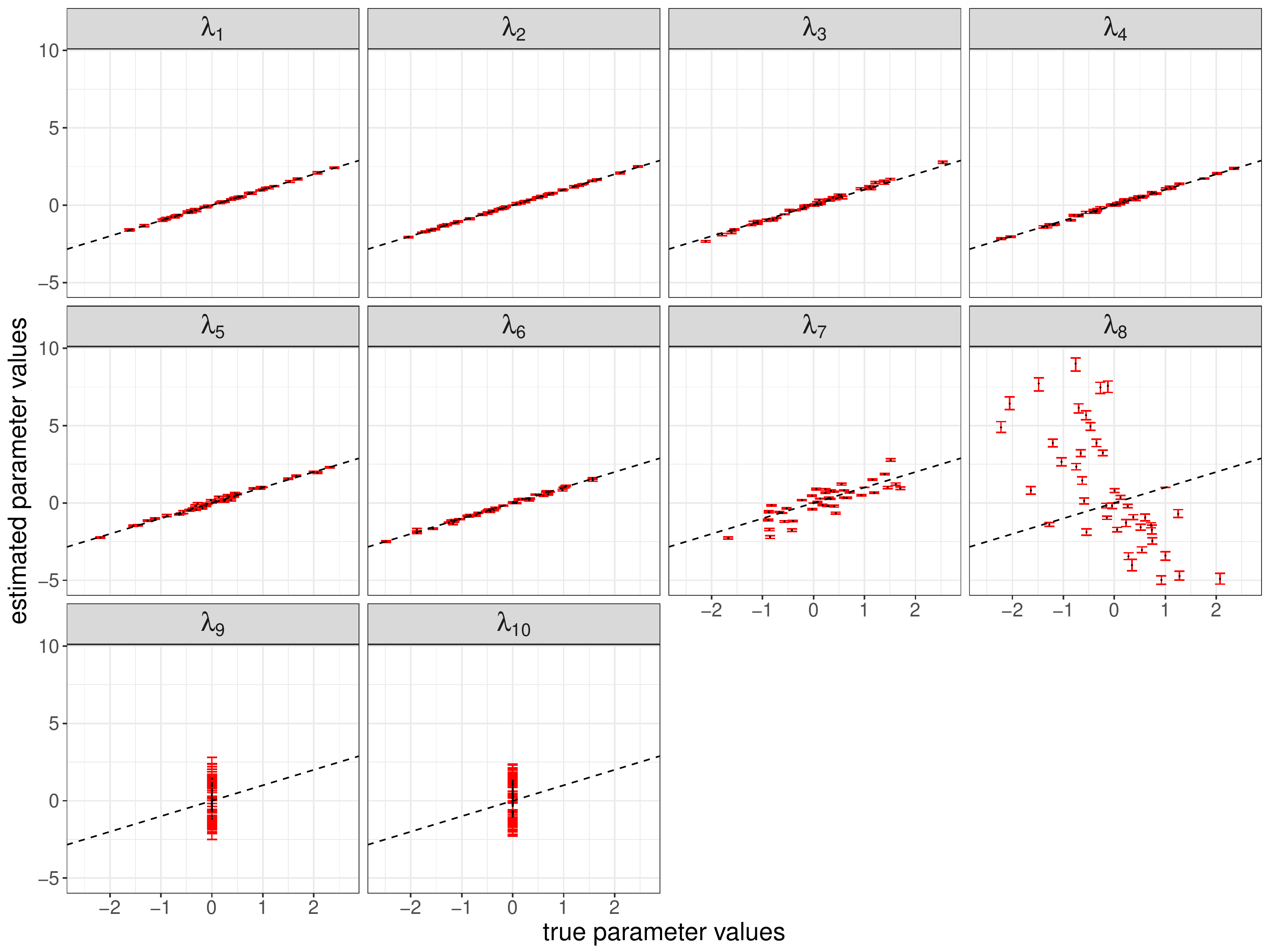}
\caption{Fitted vs true factor loadings (95\% credible sets and medians) for the model with $q_w = 10$. Each panel represents a column in $\bLambda_z$. The true parameter values for columns $\blambda_9$ and $\blambda_{10}$ are set to zero since these are not part of the true model.}\label{fig:sim1load2q10}
\end{sidewaysfigure}

\begin{figure}[ht]
\centering
\subfigure[$q_w=3$]{\includegraphics[scale=0.4, trim={0 0 0 0.7cm},clip]{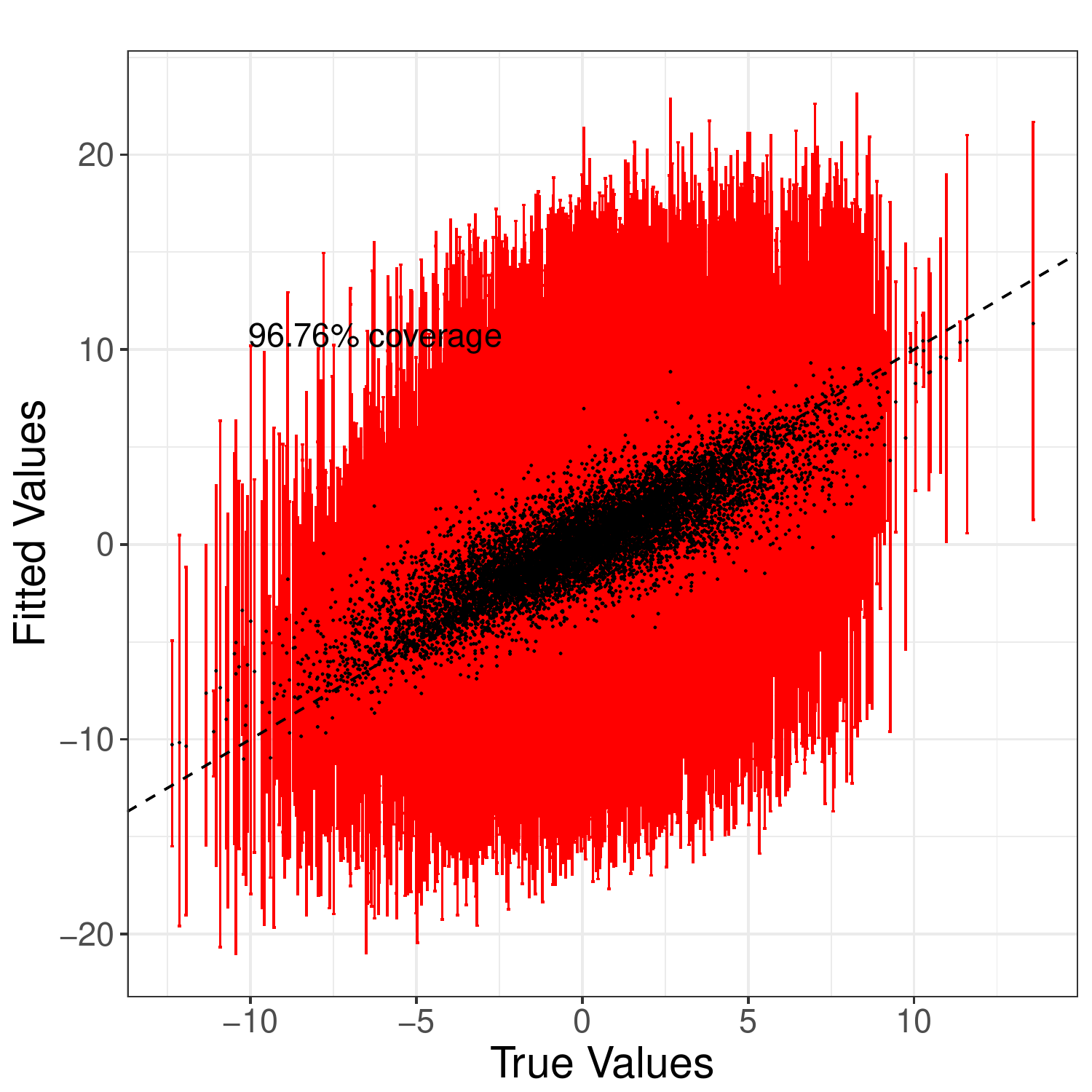}} 
\subfigure[$q_w=5$]{\includegraphics[scale=0.4, trim={0 0 0 0.7cm},clip]{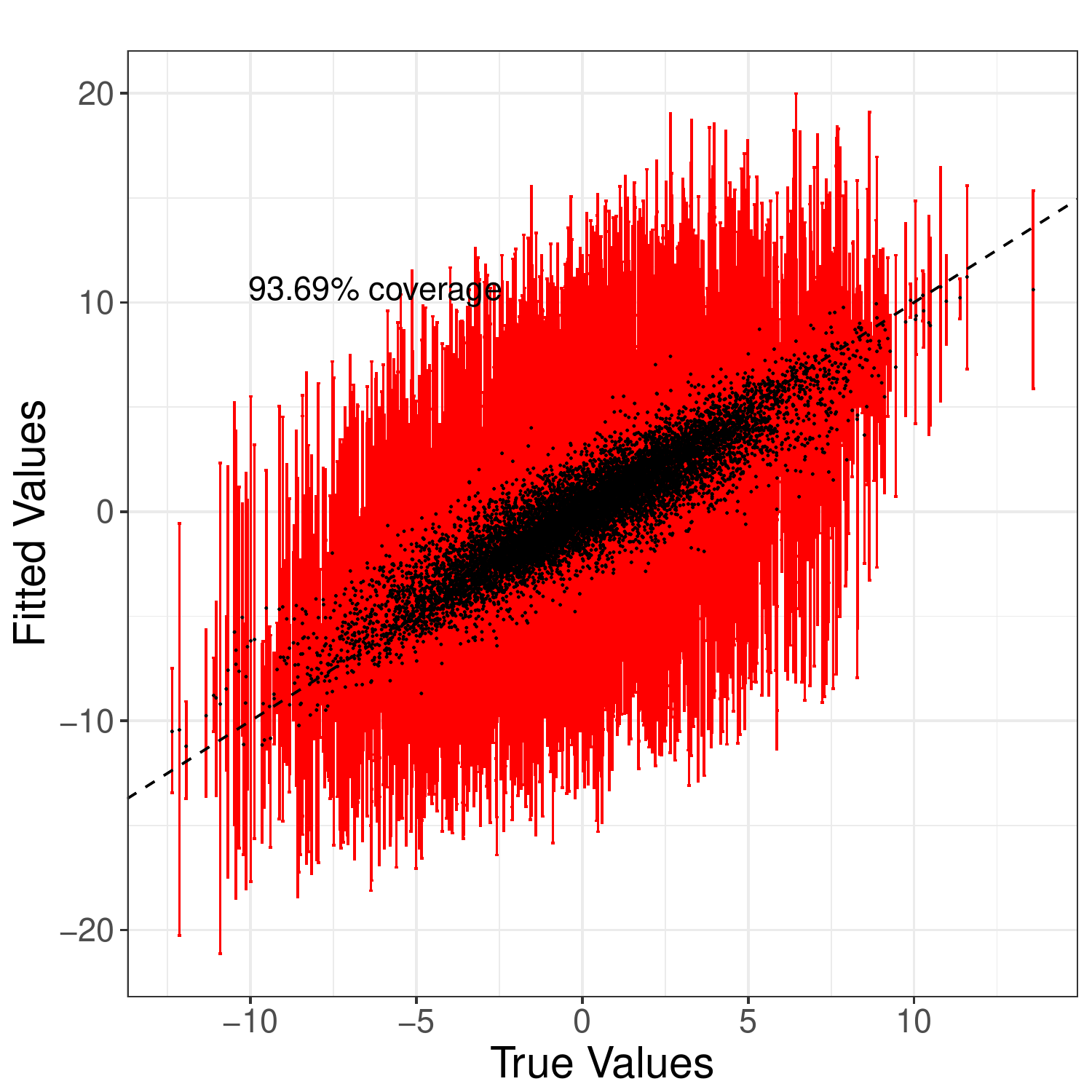}} 
\subfigure[$q_w=8$]{\includegraphics[scale=0.4, trim={0 0 0 0.6cm},clip]{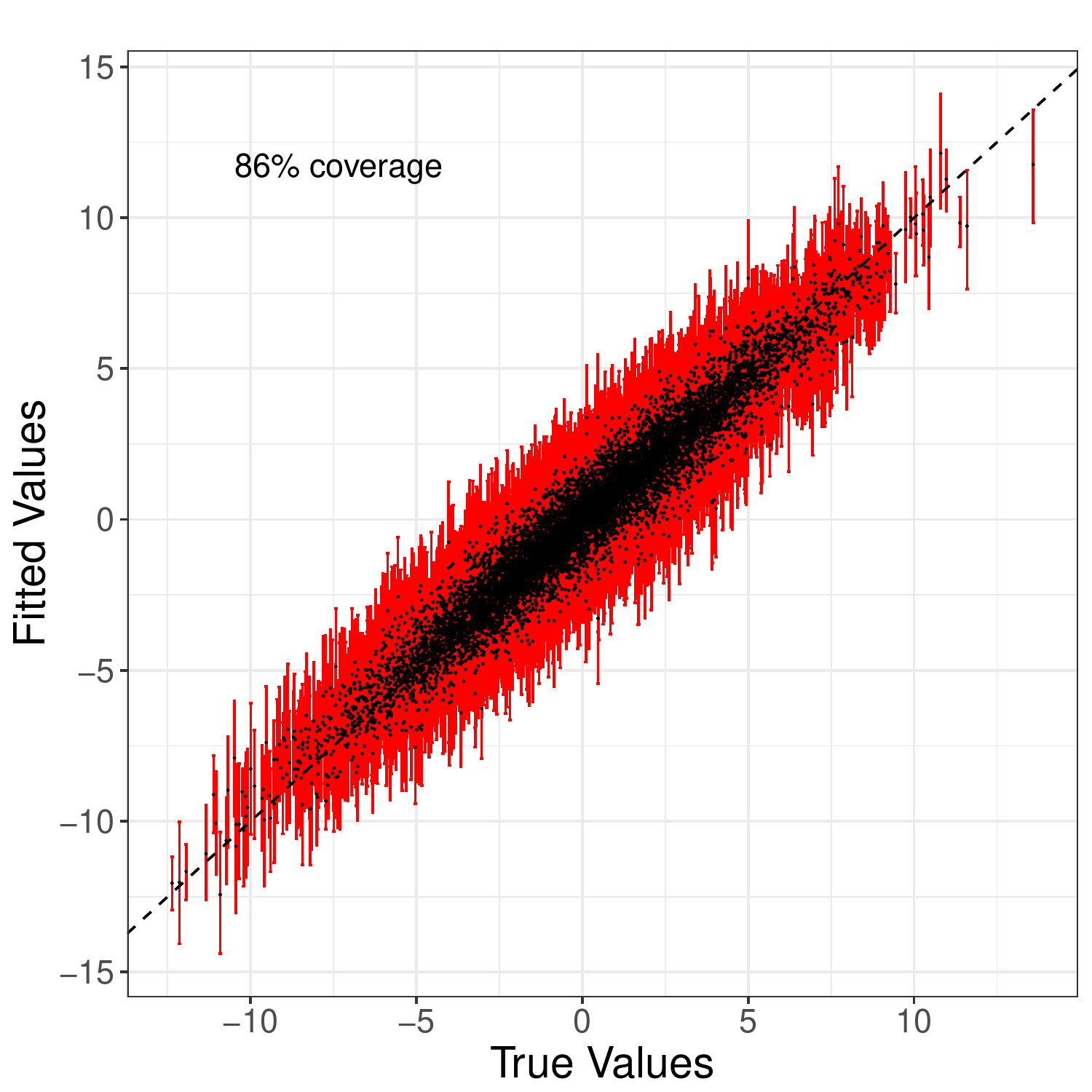}}
\subfigure[$q_w=10$]{\includegraphics[scale=0.4, trim={0 0 0 0.6cm},clip]{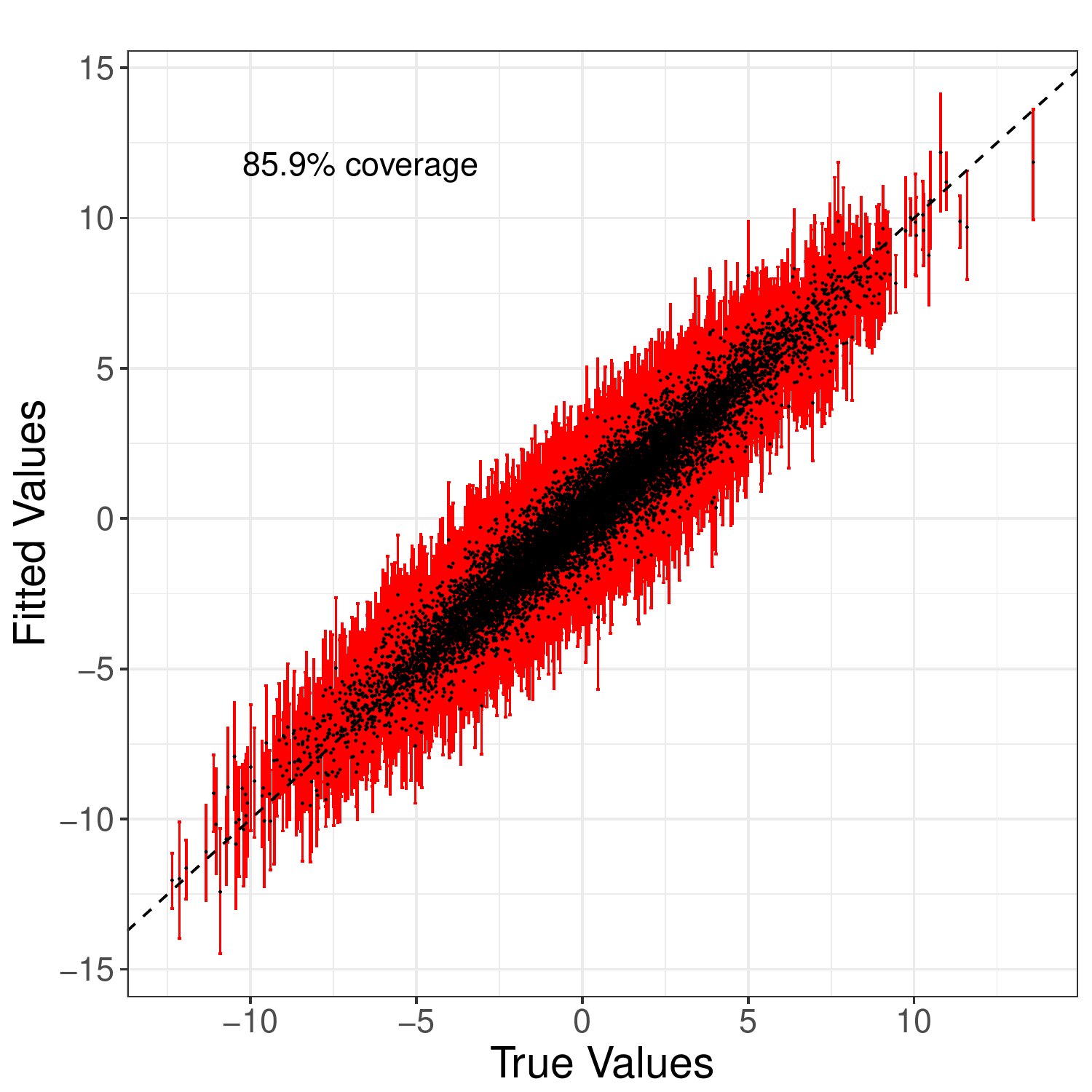}}
\caption{Imputed vs true values for missing outcomes from the simulation exercise.}\label{fig:sim1impute}
\end{figure}

\begin{figure}[ht]
\centering
\subfigure[$q_w=3$]{\includegraphics[scale=0.4, trim={0 0 0 0.7cm},clip]{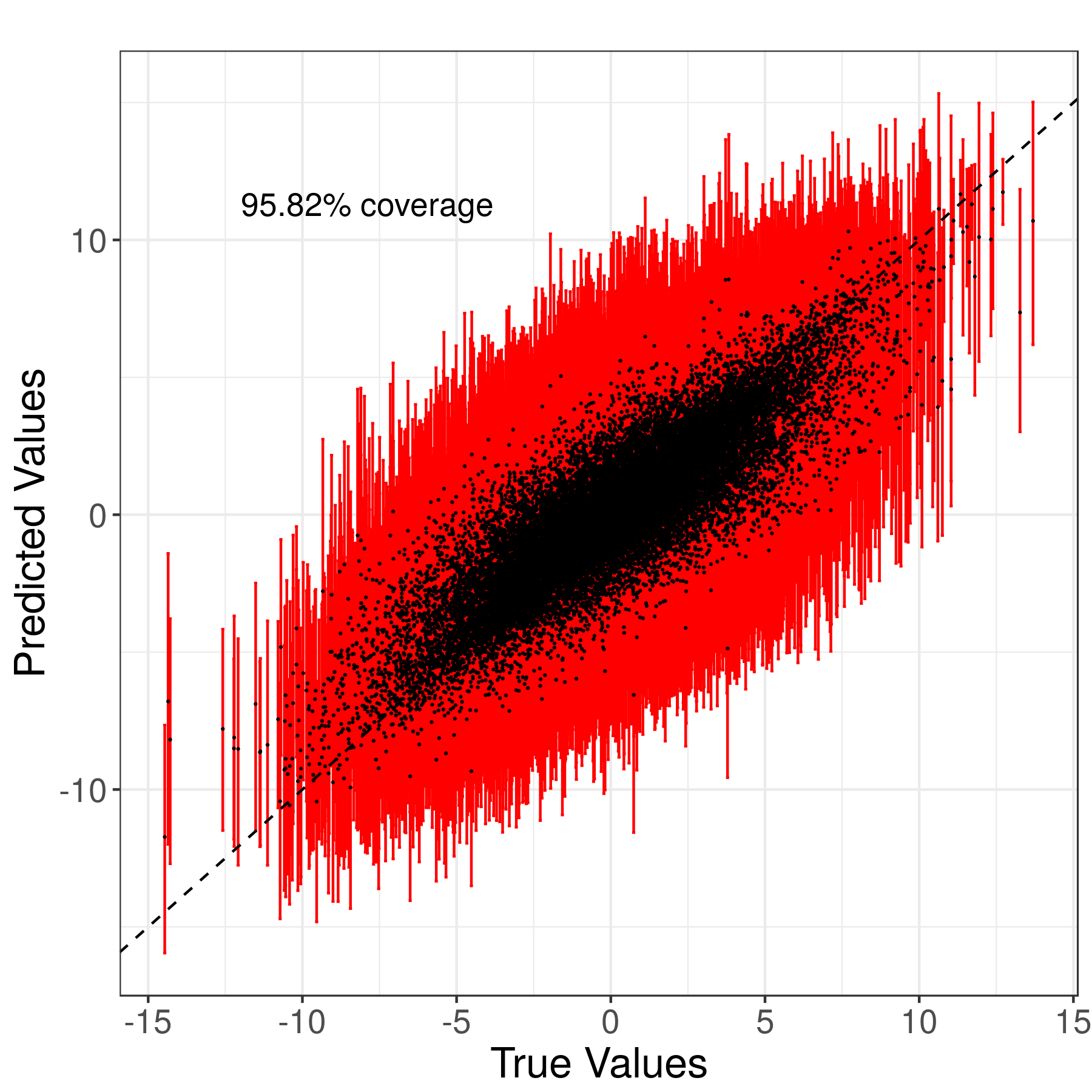}} 
\subfigure[$q_w=5$]{\includegraphics[scale=0.4, trim={0 0 0 0.7cm},clip]{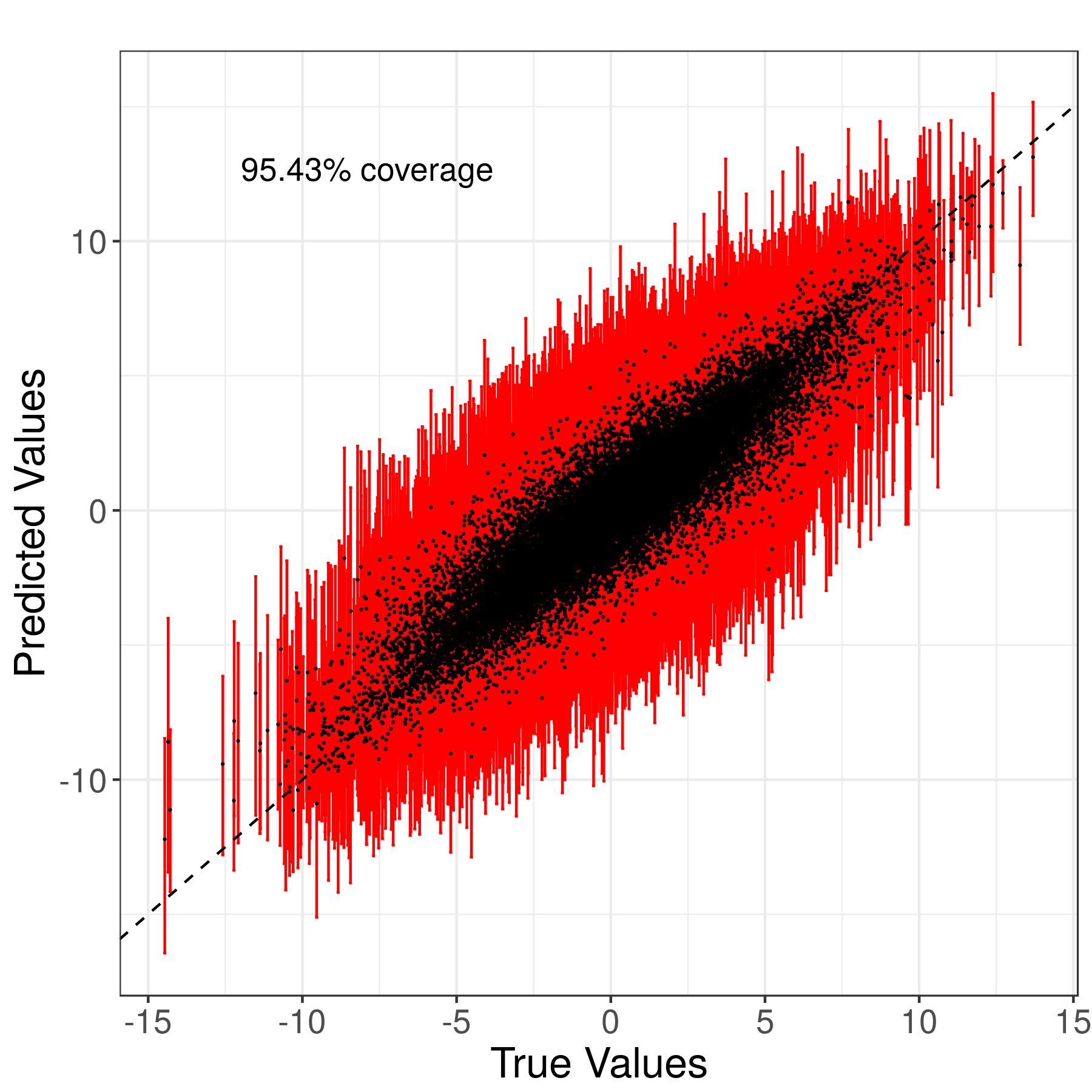}} 
\subfigure[$q_w=8$]{\includegraphics[scale=0.4, trim={0 0 0 0.6cm},clip]{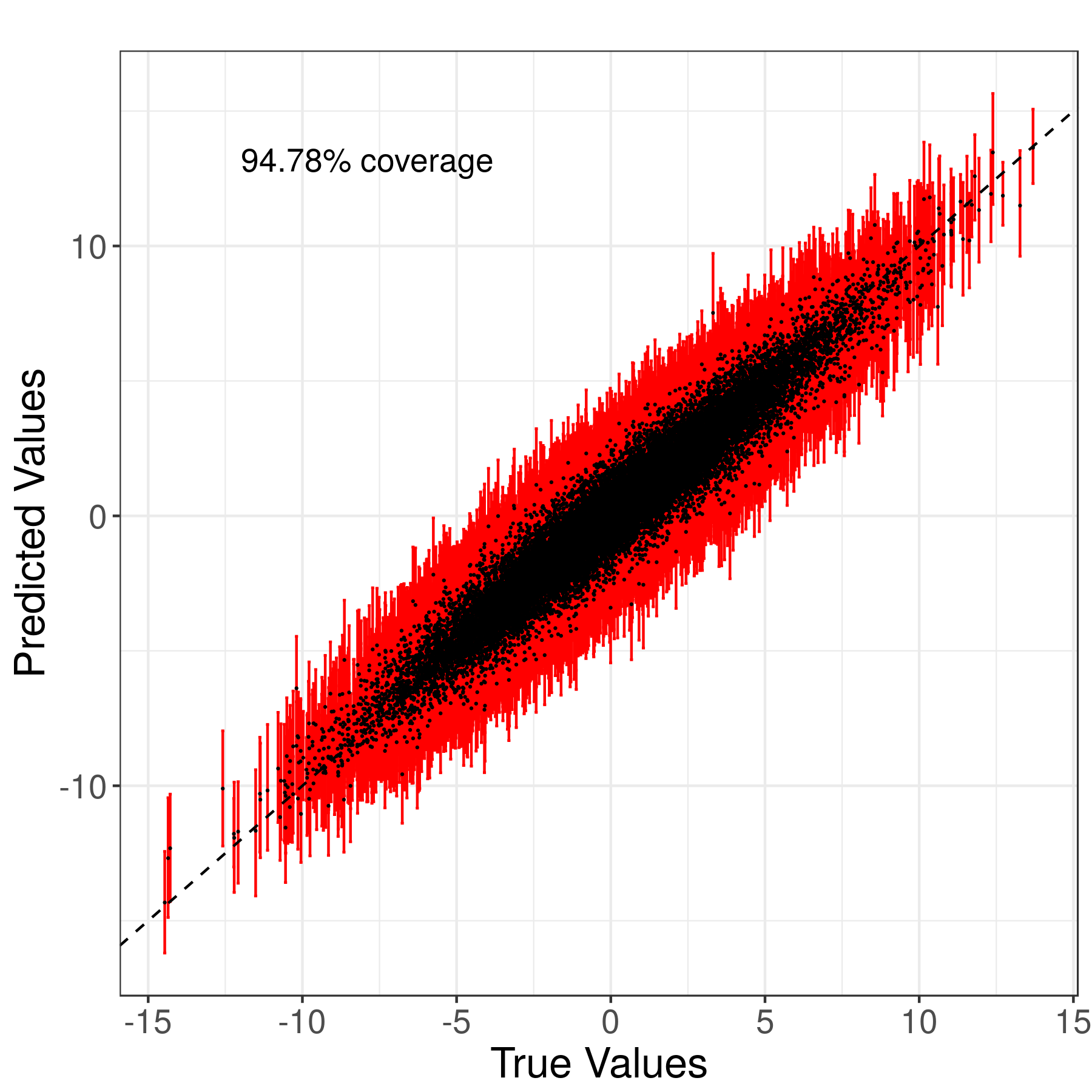}}
\subfigure[$q_w=10$]{\includegraphics[scale=0.4, trim={0 0 0 0.6cm},clip]{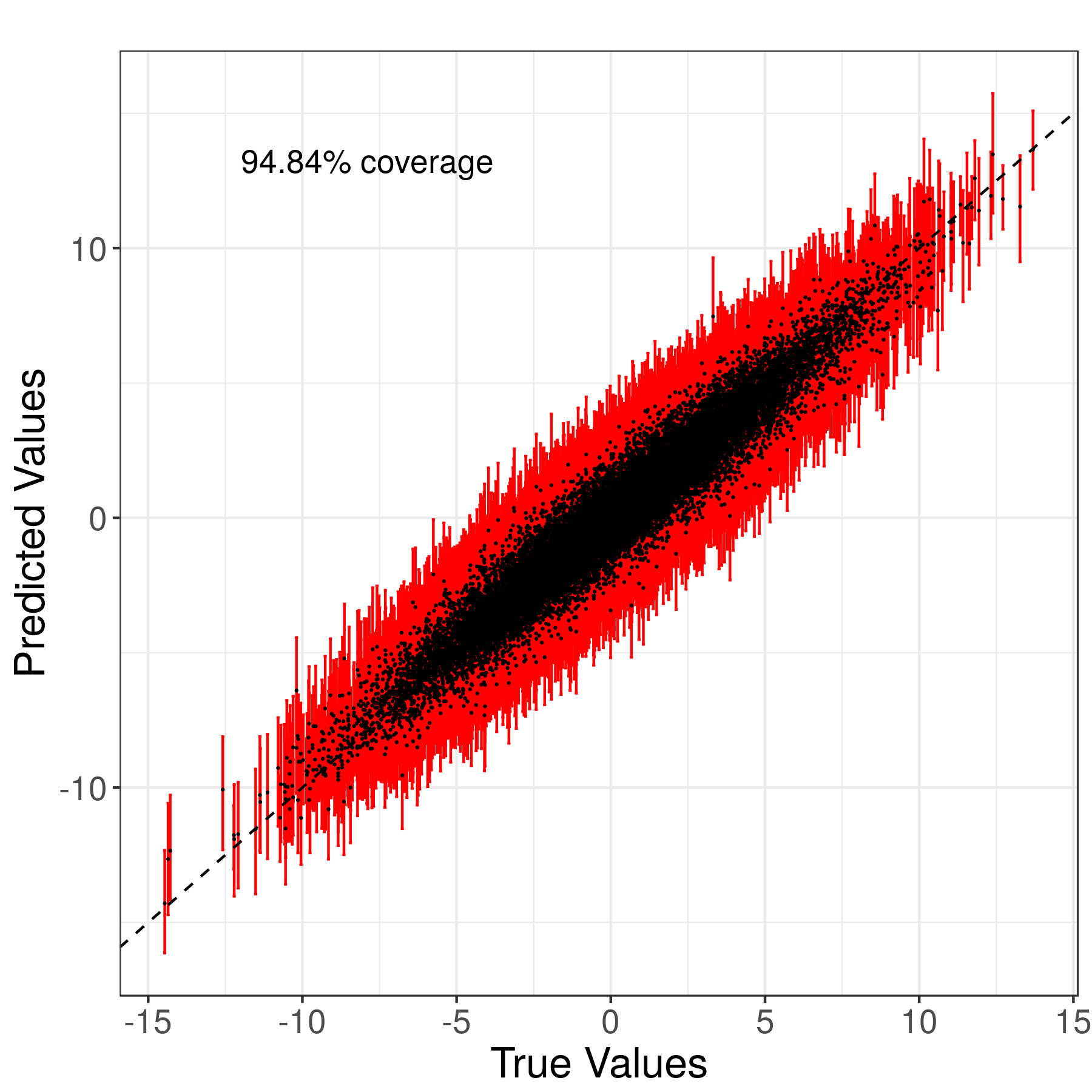}}
\caption{Predicted vs true values for 500 out-of-sample locations from the simulation exercise.}\label{fig:sim1pred}
\end{figure}



\begin{figure}[ht]
\centering
\subfigure[$q_w=3$]{\includegraphics[scale=0.27, trim={0.5cm 0.7cm 0.5cm 0},clip]{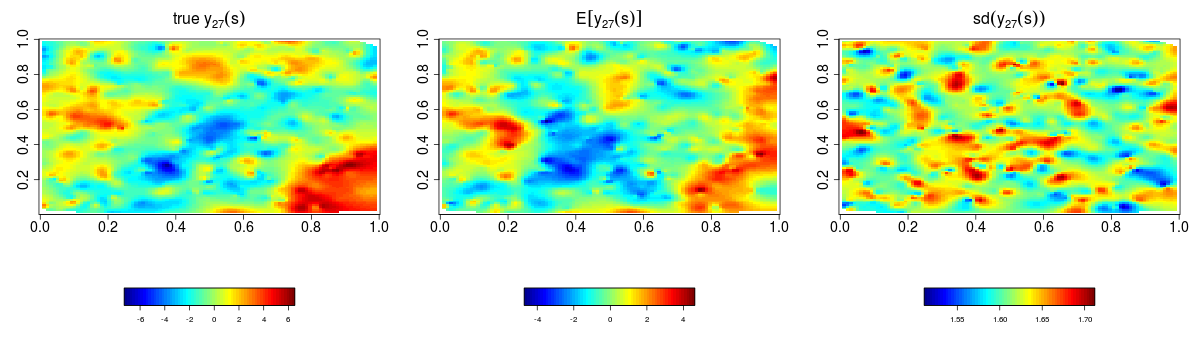}} 
\subfigure[$q_w=5$]{\includegraphics[scale=0.27, trim={0.5cm 0.7cm 0.5cm 1cm},clip]{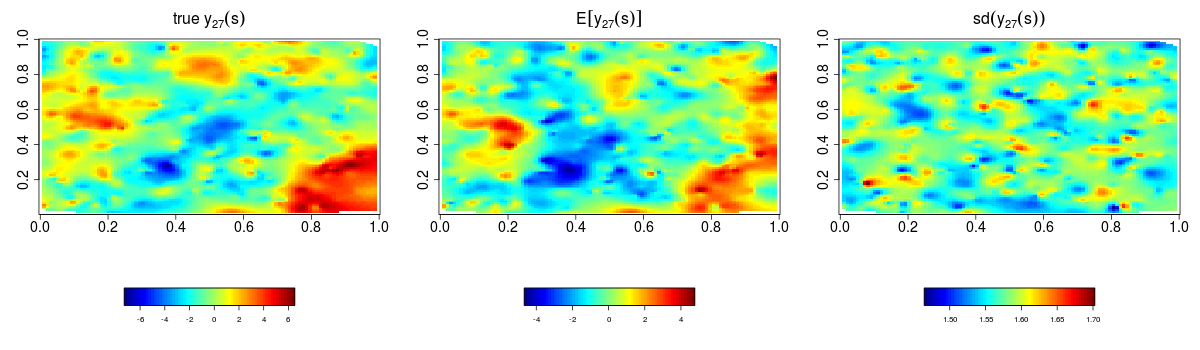}} 
\subfigure[$q_w=8$]{\includegraphics[scale=0.27, trim={0.5cm 0.7cm 0.5cm 1cm},clip]{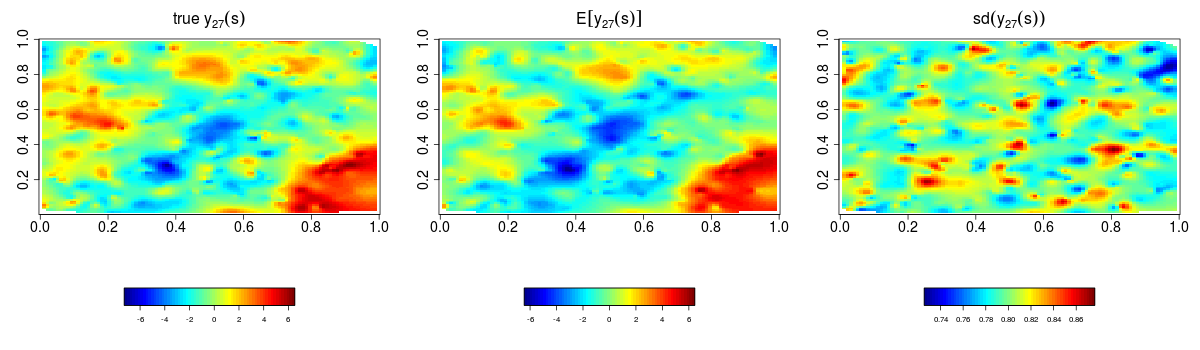}}
\subfigure[$q_w=10$]{\includegraphics[scale=0.27, trim={0.5cm 0.7cm 0.5cm 1cm},clip]{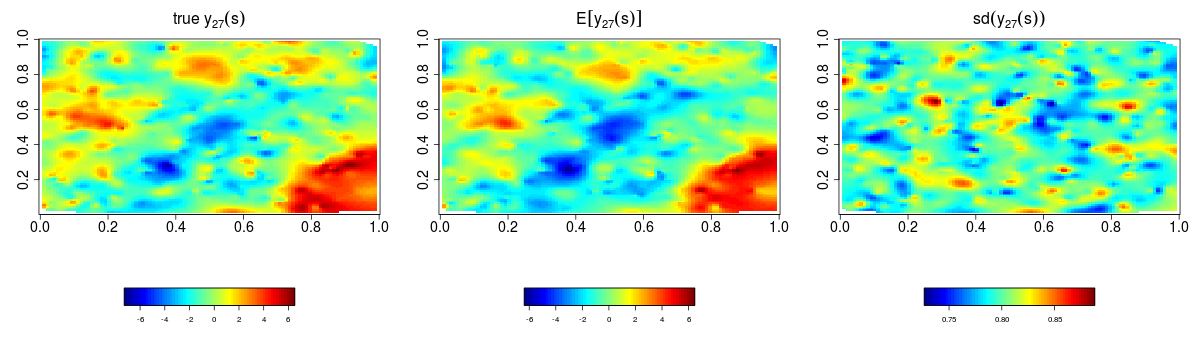}}
\caption{True, predicted and prediction uncertainty maps for $y_{27}(\cdot)$ at 500 out-of-sample locations from the simulation exercise.}\label{fig:sim1pred27}
\end{figure}
\end{document}